\documentclass[aps,prl,twocolumn,amsmath,amssymb,showpacs,superscriptaddress,notitlepage,longbibliography,floatfix]{revtex4-2}
\usepackage{graphicx}
\usepackage{subfigure}
\usepackage{dcolumn}
\usepackage{bm}
\usepackage{amsfonts}
\usepackage{mathrsfs}
\usepackage{amssymb}
\usepackage{amsmath}
\usepackage{color}
\usepackage{chngcntr}
\usepackage{stmaryrd}
\usepackage[colorlinks=true,breaklinks=true,linkcolor=red,anchorcolor=red,citecolor=red,urlcolor=red]{hyperref}
\usepackage{booktabs}
\usepackage{multirow}
\begin{document}

	\title{Half-Quantized Helical Hinge Currents in Axion Insulators}
	\date{\today}
	
\author{Ming Gong}
\affiliation{International Center for Quantum Materials, School of Physics, Peking University, Beijing 100871, China}

\author{Haiwen Liu}
\affiliation{Center for Advanced Quantum Studies, Department of Physics, Beijing Normal University, Beijing 100875, China}

\author{Hua Jiang}
\email{jianghuaphy@suda.edu.cn}
\affiliation{School of Physical Science and Technology, Soochow University, Suzhou 215006, China.}
\affiliation{Institute for Advanced Study, Soochow University, Suzhou 215006, China.}
	
\author{Chui-Zhen Chen}
\email{czchen@suda.edu.cn}
\affiliation{School of Physical Science and Technology, Soochow University, Suzhou 215006, China.}
\affiliation{Institute for Advanced Study, Soochow University, Suzhou 215006, China.}

\author{X. C. Xie}
\email{xcxie@pku.edu.cn}
\affiliation{International Center for Quantum Materials, School of Physics, Peking University, Beijing 100871, China}
\affiliation{CAS Center for Excellence in Topological Quantum Computation,
	University of Chinese Academy of Sciences, Beijing 100190, China}

\begin{abstract}
We propose that half-quantized helical hinge currents manifest as the fingerprint of the axion insulator (AI). 
These helical hinge currents microscopically originate from the lateral Goos-H\"anchen (GH) shift of massless side-surface Dirac electrons that are totally reflected from the hinges.
Meanwhile, due to the presence of the massive top and bottom surfaces of the AI, the helical current induced by the GH shift is half-quantized.
The semiclassical wave packet analysis uncovers that the hinge current has a topological origin and its half-quantization is robust to parameter changes.
Lastly, we propose an experimentally feasible six-terminal device to identify the half-quantized hinge channels by measuring the nonreciprocal conductances. Our results advance the understanding of the non-trivial transport and topological magnetoelectric responses in AIs.
\end{abstract}

	\maketitle

\emph{Introduction}.--The axion insulator (AI) is the condensed matter analog of the quasi-particle axion proposed to solve the strong $\mathrm{CP}$ problem in quantum chromodynamics \cite{peccei_cp_1977,peccei_constraints_1977,wilczek_two_1987,qi_topological_2008}. As a non-trivial topological phase, the AI is characterized by symmetry protected quantized axion coupling coefficient $\theta=\pi$ (mod $2\pi$), manifesting unique topological magnetoelectric (TME) effect \cite{qi_topological_2008,essin_magnetoelectric_2009,qi_inducing_2009,li_dynamical_2010,qi_topological_2011,sitte_topological_2012,ooguri_instability_2012,olsen_surface_2017,varnava_surfaces_2018} and has stimulated extensive research interests \cite{taherinejad_adiabatic_2015,wang_quantized_2015,wu_quantized_2016,mogi_magnetic_2017,mogi_tailoring_2017,xiao_realization_2018,allen_visualization_2019,yu_magnetic_2019,liu_robust_2020,li_nonlocal_2021,lin_direct_2021,fijalkowski_any_2021,song_delocalization_2021,li_identifying_2022,mogi_experimental_2022,zou_half-quantized_2022}.
According to the bulk-boundary correspondence of topological materials, non-trivial bulk topology induces topologically protected edge or hinge excitations in the AI \cite{xu_higher-order_2019,zhang_mobius_2020}. However, zero Hall conductance plateau has been observed in the magnetic topological insulator (TI) heterostructures \cite{mogi_magnetic_2017,mogi_tailoring_2017,xiao_realization_2018} and the antiferromagnetic TI $\rm MnBi_{2}Te_{4}$ \cite{liu_robust_2020}, which has been regarded as the characteristic of the AI with vanished total Chern number. This implies that in the AI phase, symmetry-enforced topological edge or hinge states may be absent, especially when the side surfaces inevitably enter into metallic states \cite{chu_surface_2011,chen_using_2021,gu_spectral_2021}. Whether quantized edge or hinge channels determined by the bulk-boundary correspondence theorem should exist on the metallic side surfaces of the AI, as well as their transport behaviors, are still elusive.   

To resolve the above puzzles, the microscopic mechanism of the boundary excitation in the AI should be clarified. So far, theoretical studies in the AI are mainly based on topological field theory, which cannot provide a definitive answer. Fortunately, the semiclassical analysis can give intuitive descriptions of microscopic processes in topological systems such as the adiabatic charge pumping, etc \cite{niu_quantised_1984,xiao_berry-phase_2006,xiao_berry_2010,jiang_topological_2015,PhysRevLett.115.156603,lensky_topological_2015,shi_shift_2019}. Specifically for the AI, electrons on the metallic side surfaces undergo multiple scattering between the insulating top and bottom surfaces. Therefore, it is highly desirable to illustrate the picture of these scattering processes based on semiclassical wave packet dynamics. This will further uncover the mechanism of boundary excitations in AIs and helps to provide a feasible proposal for the transport experiment to identify the AI phase.
\begin{figure}[b]
\includegraphics[width=0.49\textwidth]{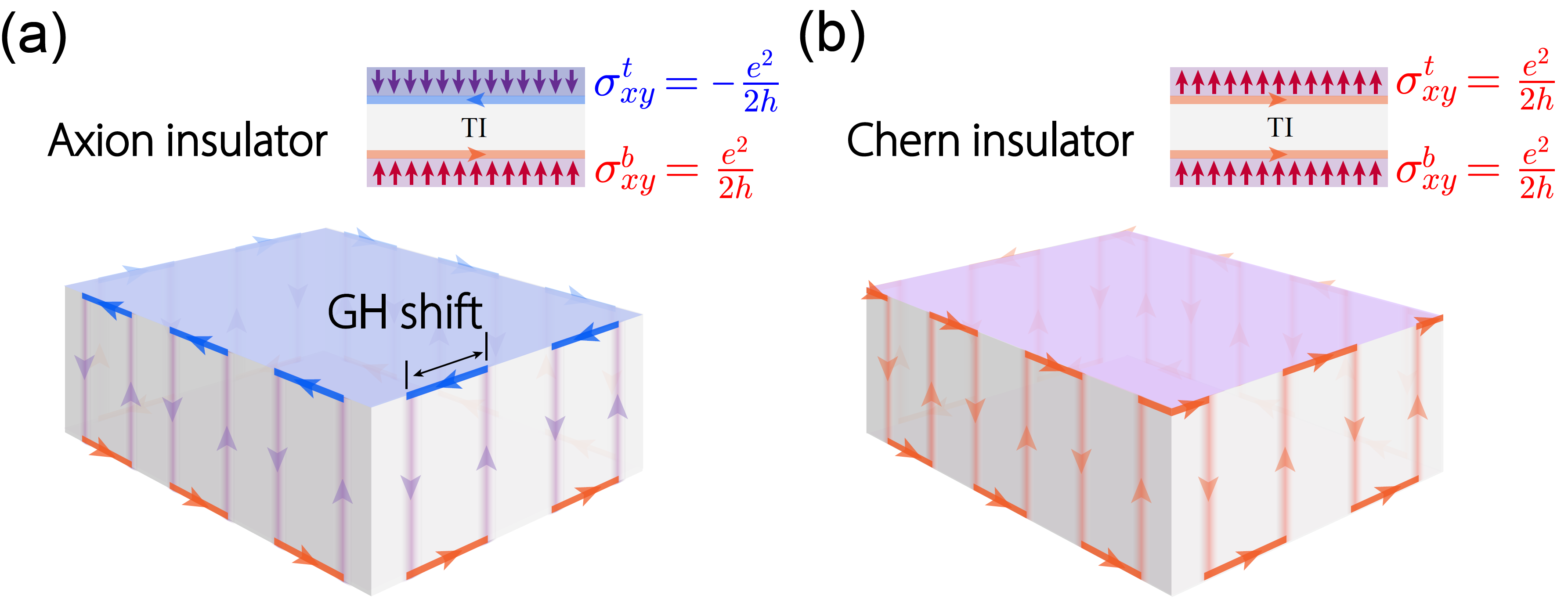}
\caption{Schematics of the half-quantized hinge currents in AIs and CIs. They originate from the GH shift currents of side-surface electrons that bounce back and forth between gapped top and bottom surfaces.}
\label{fig1}
\end{figure}

In this work, we find that half-quantized helical hinge currents exist in the AI and manifest as its fingerprint. Based on the semiclassical wave packet dynamics, we establish the microscopic picture of these hinge currents. Similar to the Goos-H\"anchen (GH) shift \cite{goos_neuer_1947,kurt_artmann_calculation_1948,renard_total_1964,yasumoto_new_1998,beenakker_quantum_2009} of the light beam, the massless Dirac electrons on the side surfaces of the AI also undergo a lateral GH shift when reflected from the massive top or bottom surface (see Fig.~\ref{fig1}). Moreover, due to the time-reversal ($\mathcal T$) symmetry-broken top and bottom surfaces, the GH shift favors a specific direction on the hinge. Consequently, helical GH shift currents are accumulated on the hinges of the AI when side-surface electrons bounce back and forth between the gapped top and bottom surfaces [see Fig.~\ref{fig1}(a)], which is distinct from the chiral net currents on the edge of the Chern insulator (CI) [see Fig.~\ref{fig1}(b)]. Interestingly, we find that the differential GH shift current $\delta I_{GH}$ is exactly half-quantized with respect to the differential Fermi energy $\delta E_{F}$, i.e. $\delta I_{GH}=e/2h\delta E_{F}$.
We demonstrate that the GH shift current originates from the non-vanishing Berry curvature during the scattering process, and its half-quantization is robust to the variation of parameters.
Then, we numerically verify the half-quantized helical hinge currents
through a 3D lattice model.
Finally, we propose a six-terminal device to identify the half-quantized hinge currents in the AI by measuring the nonreciprocal conductances.

\begin{figure}[t]
\includegraphics[width=0.49\textwidth]{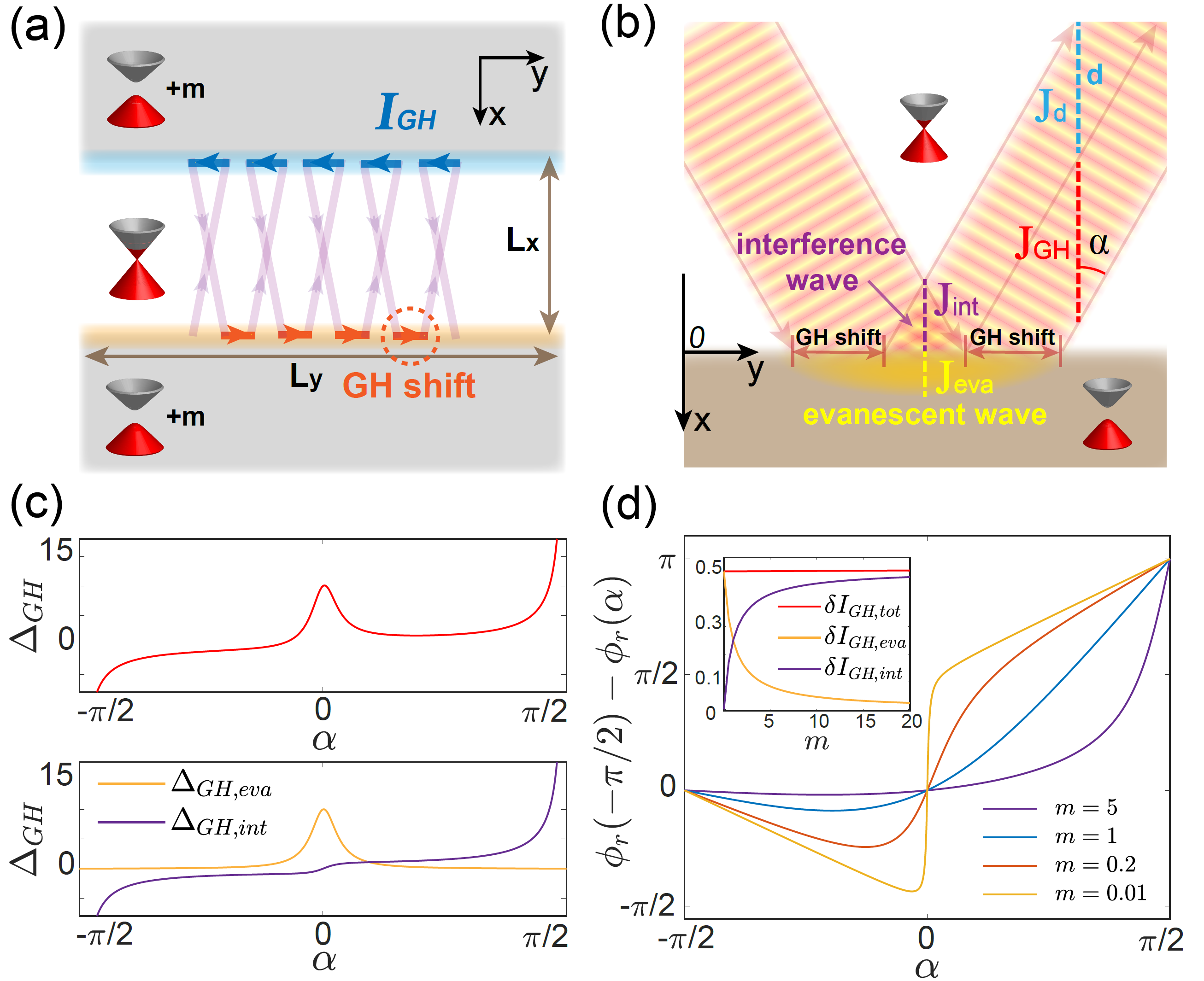}
\caption{(a) Unfolded view of the top-side-bottom surfaces of the AI. The trajectory of a massless electron illustrates the formation of the GH shift current $I_{GH}$. When the contributions from all the electrons are considered, the trajectory segments on the hinges are merged together, leading to net helical hinge currents. (b) Sketch of the scattering process on the hinge, where a beam of massless Dirac electron is totally reflected from a massive barrier. (c) The GH shift versus $\alpha$ with $U=1$, $m=0.1$, and $E=0$. The upper panel shows the total GH shift $\Delta_{GH}$ and the lower panel shows its contributions from the evanescent wave ($\Delta_{GH,eva}$) and the interference wave ($\Delta_{GH,int}$). (d) Reflection phase difference versus $\alpha$ for different $m$. The inset figure shows the evanescent wave and interference wave contributions to $\delta I_{GH}$ (in units of $e/h\delta E_{F}$) versus $m$.}
\label{fig2}
\end{figure}

\emph{Model Hamiltonian and GH shift}.--We model the AI by a 3D magnetic TI, of which the top and bottom surfaces are gapped oppositely while the side surfaces remain gapless. Such a model is in accordance with the experimental setups utilized to realize the AI state in magnetic TI heterostructure or antiferromagnetic TI $\rm MnBi_{2}Te_{4}$ \cite{mogi_magnetic_2017,mogi_tailoring_2017,xiao_realization_2018,liu_robust_2020}. According to the bulk-boundary correspondence theorem, the top/bottom surface and the side surface of the AI can be described by the 2D effective Hamiltonians:
	\begin{equation}\label{eq:1}
	\mathcal H=\left\{
	\begin{array}{ll}
	\hbar v_{F}(-i\sigma_{x}\partial_{x}-i\sigma_{y}\partial_{y})-U&({\rm side})\\
	\hbar v_{F}(-i\sigma_{x}\partial_{x}-i\sigma_{y}\partial_{y})+m\sigma_{z}&({\rm top/bottom})
	\end{array}\right..
	\end{equation}
Here, $v_{F}$ denotes the Fermi velocity, $U$ is the gate potential, and $m$ is the mass term induced by the magnetization. 
In the following discussions, we unfold the top-side-bottom surfaces of the AI [see Fig.~\ref{fig1}(a)] into the $x$-$y$ plane [see Fig.~\ref{fig2}(a)].
	
We use the probability flux method \cite{renard_total_1964,yasumoto_new_1998} to calculate the GH shift on the hinge of the AI. As shown in Fig.~\ref{fig2}(b), we place the massless side surface in $x\leqslant0$ and the massive top/bottom surface in $x>0$.
Firstly, we solve the scattering problem by matching $\psi(\bm{r})=\psi_{\rm{in}}+r\psi_{\rm{re}}$ ($x\leqslant 0$) and $\psi_{{\rm eva}}(\bm{r})=e^{-\kappa x+ik_{y}y}\psi(0,0)$ ($x>0$) at $x=0$, where $\psi(\bm{r})$ is the superposition of the incident plane wave $\psi_{\rm{in}}(\bm{r})=e^{ik_{x}x+ik_{y}y}[e^{-i\frac{\alpha}{2}},e^{i\frac{\alpha}{2}}]^{T}/\sqrt{2}$ and reflected plane wave $\psi_{\rm{re}}(\bm{r})=e^{-ik_{x}x+ik_{y}y}[e^{-i\frac{\pi-\alpha}{2}},e^{i\frac{\pi-\alpha}{2}}]^{T}/\sqrt{2}$ in the massless region, and $\psi_{{\rm eva}}(\bm{r})$ is the evanescent wave in the massive region. $\alpha={\rm arctan}\frac{k_{y}}{k_{x}}$ denotes the incident angle [see Fig.~\ref{fig2} (a)] and $\bm{r}=(x,y)$. Here we only consider the total reflection process, which captures the physics within the magnetization gap. The reflection coefficient $r=e^{i\phi_{r}}$ is given in \cite{Note1}. 
Then, imagine that the incident and reflected waves have finite width as depicted in Fig.~\ref{fig2}(b), the GH shift can be obtained under the probability flux conservation constraint. As labeled in Fig.~\ref{fig2}(b), $J_{\rm{int}}$ ($J_{\rm{eva}}$) represents the flux carried by the interference wave (evanescent wave), $J_{\rm{GH}}$ is the flux induced by GH shift, while $J_{\rm{d}}$ is the flux through the cross-section colored blue with width $d$. The flux conservation condition imposes that $J_{\rm{int}}+J_{\rm{eva}}=J_{\rm{GH}}+J_{\rm{d}}$. Following the standard flux method analysis \cite{renard_total_1964,yasumoto_new_1998,Note1}, we obtain
\begin{eqnarray}
\Delta_{GH}&&=\frac{{\rm sin}\alpha+{\rm cos}\phi_{r}}{\kappa {\rm cos}\alpha}-\frac{{\rm sin}\phi_{r}}{k_{x}{\rm cos}\alpha}.
\label{eq:2}
\end{eqnarray}

In the upper panel of Fig.~\ref{fig2}(c), we plot $\Delta_{GH}$ as a function of the incident angle $\alpha$ with fixed $E$, $U$, and $m$. For glancing incidence, i.e. $\alpha \to \pm\pi/2$, $\Delta_{GH}$ diverges but with opposite sign for $\alpha=\pi/2$ and $\alpha=-\pi/2$. Such behavior indicates that at large incident angles, the GH shift has the same direction as the incident wave, thus does not show the chiral feature. However, $\Delta_{GH}$ peaks for vertical incidence, i.e. $\alpha \to 0$, showing clearly that the GH shift favors a specific direction and manifests the chiral feature.  The chiral $\Delta_{GH}$ further implies that the net GH shift current accumulated on the hinge is chiral.  
We then decompose $\Delta_{GH}$ according to its contributions from the evanescent wave $\Delta_{GH,eva}$ and the interference wave $\Delta_{GH,int}$ \cite{Note1}. As will be discussed later, $\Delta_{GH,eva}$ and $\Delta_{GH,int}$ introduce the components of the shift current with different decay laws. The lower panel of Fig.~\ref{fig2}(c) shows that the chiral part of $\Delta_{GH}$ is mostly carried by $\Delta_{GH,eva}$, which lies on the $\mathcal T$-symmetry-broken top/bottom surface of the AI. In contrast, the non-chiral part, especially for glancing incidence $\alpha\rightarrow\pm \pi/2$, is mostly carried by $\Delta_{GH,int}$.

\emph{half-quantized GH shift current}.--
To give a microscopic picture of the chiral hinge current induced by the GH shift, we consider the electron in the scattering problem as a point particle, which bounces back and forth between double massive barriers as sketched in Fig.~\ref{fig2}(a). Suppose the width between the opposite hinges is $L_{x}$, thus the average time interval between two consecutive bounces is $\Delta \tau =2L_{x}/v_{F}\rm{cos}\alpha$ for $-\pi/2<\alpha <\pi/2$. When it bounces off the hinge, it undergoes a lateral GH shift $\Delta_{GH}$. Such a lateral shift induces an anomalous velocity of electrons along the hinge as
\begin{eqnarray}
v_{GH}=\frac{\Delta_{GH}}{\Delta\tau}=\Delta_{GH}v_{F}{\rm cos}\alpha/2L_{x}.
\end{eqnarray} 
The total GH shift current induced by the $v_{GH}$ is obtained by counting the contributions of all filled electrons as 
\begin{eqnarray}
I_{GH}&&=\sum_{\rm filled} \frac{ev_{GH}}{L_{y}}=\frac{e}{h}\int_{-\infty}^{E_{F}}dE\int_{-K}^{K}\frac{dk_{y}}{2\pi}\Delta_{GH},
\end{eqnarray}
where $L_{y}$ is the circumference of the side surface. $K=(E+U)/\hbar v_{F}$ is the Fermi wave vector at energy $E$. Details for calculations can be found in \cite{Note1}.
By employing the stationary phase method \cite{kurt_artmann_calculation_1948,Note1}, we find that the GH shift can be written as $\Delta_{GH}=-\partial \phi_{r}/\partial k_{y}$.
Plugging this result into Eq.~(4), we obtain the differential shift current
$\delta I_{GH}=\delta E_{F}[\phi_{r}(-\pi/2)-\phi_{r}(\pi/2)]e/2\pi h$.

In Fig.~\ref{fig2}(d), we plot $\phi_{r}$ as a function of $\alpha$  for different $m$. It is clear that $\phi_{r}(-\pi/2)-\phi_{r}(\pi/2)=\pi$ and robust to the variation of $m$ \footnote{see Supplementary Materials, which includes Refs. \cite{marchand_lattice_2012,thouless_quantized_1982,thouless_quantization_1983,lu_competition_2011,zhang_non-reciprocal_2022} \label{sup}}. Therefore, $\delta I_{GH}$ is exactly half-quantized with respect to $\delta E_{F}$ as 
\begin{eqnarray}
\delta I_{GH}=\frac{e}{2h}\delta E_{F}.
\end{eqnarray}
Moreover, $\delta I_{GH}$ can be decomposed according to its contributions from the evanescent wave and the interference wave, i.e. $\delta I_{GH}=\delta I_{GH,eva}+\delta I_{GH,int}$. We prove in \cite{Note1} that $\delta I_{GH,int}$ is power low ($x^{-\frac12}$) decay from the hinge, while $\delta I_{GH,eva}$ decays exponentially from the hinge ($e^{-x/\lambda}$). Therefore, $\delta I_{GH}$ is different from the current carried by the topologically protected edge or hinge state, which decays exponentially on both sides of the hinge. The inset figure of Fig.~\ref{fig2}(d) plots $\delta I_{GH,eva}$ and $\delta I_{GH,int}$ versus $m$. $\delta I_{GH}$ is mainly contributed from $\delta I_{GH,eva}$ when $m$ is small. 
As $m$ increases, the contribution from $\delta I_{GH,int}$ becomes dominant.

\emph{Topological origin of the GH shift current.}--
We provide a topological viewpoint of the half-quantized GH shift current in the frame of the adiabatic charge transport theory \cite{niu_quantised_1984,xiao_berry_2010}. Here, we use Hamiltonian $\mathcal H(\bm{r})=\hbar v_{F}(-i\sigma_{x}\partial_{x}-i\sigma_{y}\partial_{y})+m(x)\sigma_{z}$ to describe the scattering process,
where $m(x)$ is a smooth function connecting the gapless and the gapped regions with $m(x)\to m$ for $x\gg 0$ and $m(x)\to 0$ for $x\ll 0$.
During the reflection process, the energy $E$ and momentum $k_{y}$ are conserved. Therefore, the relation $\hbar^{2}v_{F}^{2}k_{x}^{2}+\hbar^{2}v_{F}^{2}k_{y}^{2}+m^{2}=E^{2}$ holds.
We take $\hbar v_{F} k_{x}= -t$ as the virtual time and the scattering process is now described by the time-dependent 1D Hamiltonian
$H(k_{y},t)=-t\sigma_{x}+\hbar v_{F} k_{y}\sigma_{y}+m(t)\sigma_{z}$.
The study of the GH shift is reduced to the study of an adiabatic charge transport problem in the $y$-direction. Define the Berry curvature $\Omega_{k_{y}t}=-2{\rm Im} \langle \partial u/\partial k_{y}|\partial u/\partial t \rangle$,
where $|u(k_{y},t)\rangle$ represents the instantaneous eigenstate of $H(k_{y},t)$. According to the adiabatic charge transport theory, $v_{y}(k_{y})=\partial E(k_{y})/\hbar\partial k_{y}-\Omega_{k_{y}t}$ is the velocity of the electron in the $y$-direction. The GH shift in the $y$-direction for a given $k_{y}$ is
\begin{eqnarray}
\Delta_{GH}(k_{y})=\int_{-T}^{T}v_{y}dt= \int_{-T}^{T}\left[ \frac{\partial E(k_{y})}{\hbar\partial k_{y}}-\Omega_{k_{y}t} \right]dt
\end{eqnarray} 
with $T=\sqrt{E^{2}-\hbar^{2}v_{F}^{2}k_{y}^{2}}$. Combine it with Eq.~(4), one obtains
\begin{eqnarray}
\frac{\delta I_{GH}}{\delta E_{F}}&&=-\frac{e}{h}\int_{-\frac{E}{\hbar v_{F}}}^{\frac{E}{\hbar v_{F}}} \frac{dk_{y}}{2\pi} \int_{-T}^{T}\Omega_{k_{y}t} dt =-\frac{e}{2\pi h}\Gamma(C).
\end{eqnarray} 
The integration of $\partial E(k_{y})/\hbar\partial k_{y}$ vanishes because the band structure is symmetric with respect to $k_{y}$.
Note that Eq.~(7) is nothing but the Berry phase along the boundary $C$ of the integration manifold. As we have proved in \cite{Note1}, $\Gamma(C)=-{\rm sgn}(m)\pi$, which is determined by the $\pi$ Berry phase of massless Dirac electron and the sign of the massive barrier, thus is robust to local parameter changes.

\begin{figure}[t]
\includegraphics[width=0.49\textwidth]{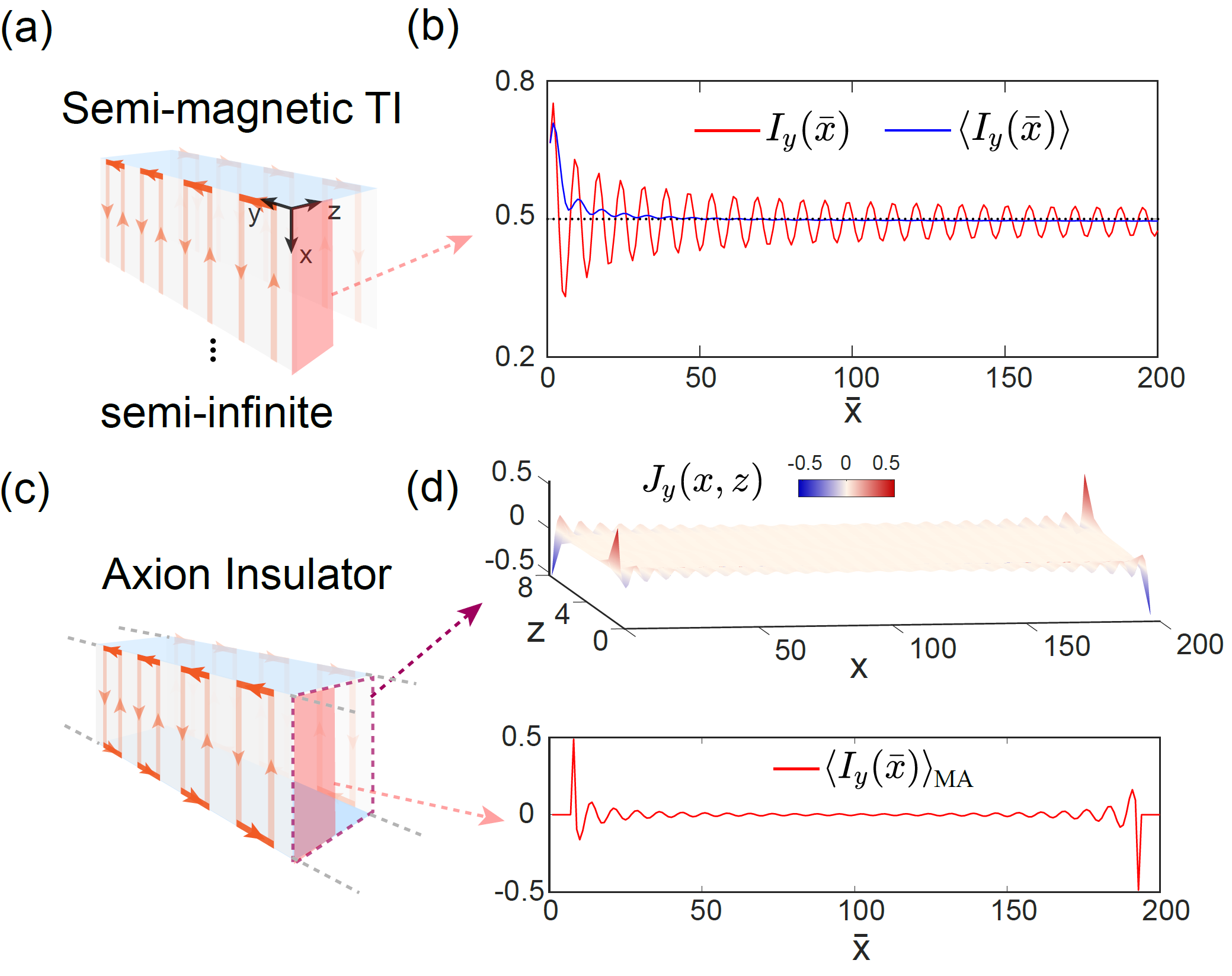}
\caption{Current density distribution in the $x$-$z$ plane for semi-magnetic TI and AI. (a) and (c), schematics of semi-magnetic TI and AI (infinite in the $y$-direction). The thickness in the $z$-direction is $L_{z}=8$. The pinked areas in (a) and (c) show the region ($0\leqslant z\leqslant L_{z}/2$) and the dotted box in (c) shows the cross-section in $x$-$z$ plane. (b) The current flux through region $[0\leqslant x\leqslant \bar{x}, 0\leqslant z\leqslant L_{z}/2]$ and $\bar{x}$ averaged $\langle I_{y}(\bar{x}) \rangle$ for semi-magnetic TI. (d) The upper panel shows the distribution of $J_{y}(x,z)$ in the $x$-$z$ plane. The lower panel shows the moving averaged current flux $\langle I_{y}(\bar{x})\rangle_{\rm MA}$ through the window $[\bar{x}-7\leqslant x\leqslant \bar{x}+7, 0\leqslant z\leqslant L_{z}/2]$. Both $J_{y}$ and $I_{y}$ are in units of $e/h$.}
\label{fig3}
\end{figure}
\emph{Current density distribution on the hinges.}--
The half-quantized hinge current can be numerically visualized based on the 3D magnetic TI Hamiltonian. The model Hamiltonian is given by
$H = H_{0}+H_{M}$, where $H_{0}=\sum_{i=x,y,z} Ak_{i}\tau_{x}\otimes \sigma_{i}+(M_{0}-Bk^{2})\tau_{z}\otimes\sigma_{0}$ is the nonmagnetic part and $H_{M}=M(\bm{r})\tau_{0}\otimes\sigma_{z}$ represents the magnetization term \cite{liu_model_2010,zhang_topological_2009}.
Here, $\sigma_{i}$ and $\tau_{i}$ are the Pauli matrices acting on the spin and orbital spaces respectively.
We use formula 
\begin{eqnarray}
J_{y}(E,\bm{r})=-\frac{e}{\pi h}\int_{-\pi}^{\pi}{\rm ImTr}[\frac{\partial H(k_{y})}{\partial k_{y}}G^{r}_{k_{y}}(E,\bm{r},\bm{r})]dk_{y}
\end{eqnarray}
\cite{chu_surface_2011} to obtain the current density in the $y$-direction across the $x$-$z$ plane with $\bm{r}=(x,z)$. Here, $G^{r}_{k_{y}}(E,\bm{r},\bm{r'})$ is the retarded Green's function for the momentum-sliced Hamiltonian $H(k_{y})$.

We first evaluate a semi-magnetic TI with a gapped top surface in $y$-$z$ plane, which can be viewed as a ``half" of an AI [see Fig.~\ref{fig3}(a)] \cite{mogi_experimental_2022,zhou_transport_2022,zou_half-quantized_2022}.
As shown in Fig.~\ref{fig3}(b), the current flux $I_{y}(\bar x)=\int_{0}^{L_{z}/2}dz\int_{0}^{\bar x}dx J_{y}(x,z)$ oscillates around 0.5. The $\bar x$ averaged $\langle I_{y}(\bar x) \rangle=\int_{0}^{\bar x}I_{y}(\bar x')d\bar x'/\bar x \to 0.5$ clearly shows the half-quantization of the chiral hinge current, which coincides with the GH shift analysis.
For an AI where the top and bottom surfaces are gapped with opposite magnetization [see Fig.~\ref{fig3}(c) and (d)], a pair of counter-propagating hinge currents emerges on the hinges. The moving averaged current flux $\langle I_{y}(\bar{x})\rangle_{\rm MA}=\int_{0}^{L_{z}/2}dz\int_{\bar x-7}^{\bar x+7}dxJ_{y}(x,z)$ further demonstrates that the hinge currents are nearly half-quantized and helical.

\begin{figure}[t]
\includegraphics[width=0.49\textwidth]{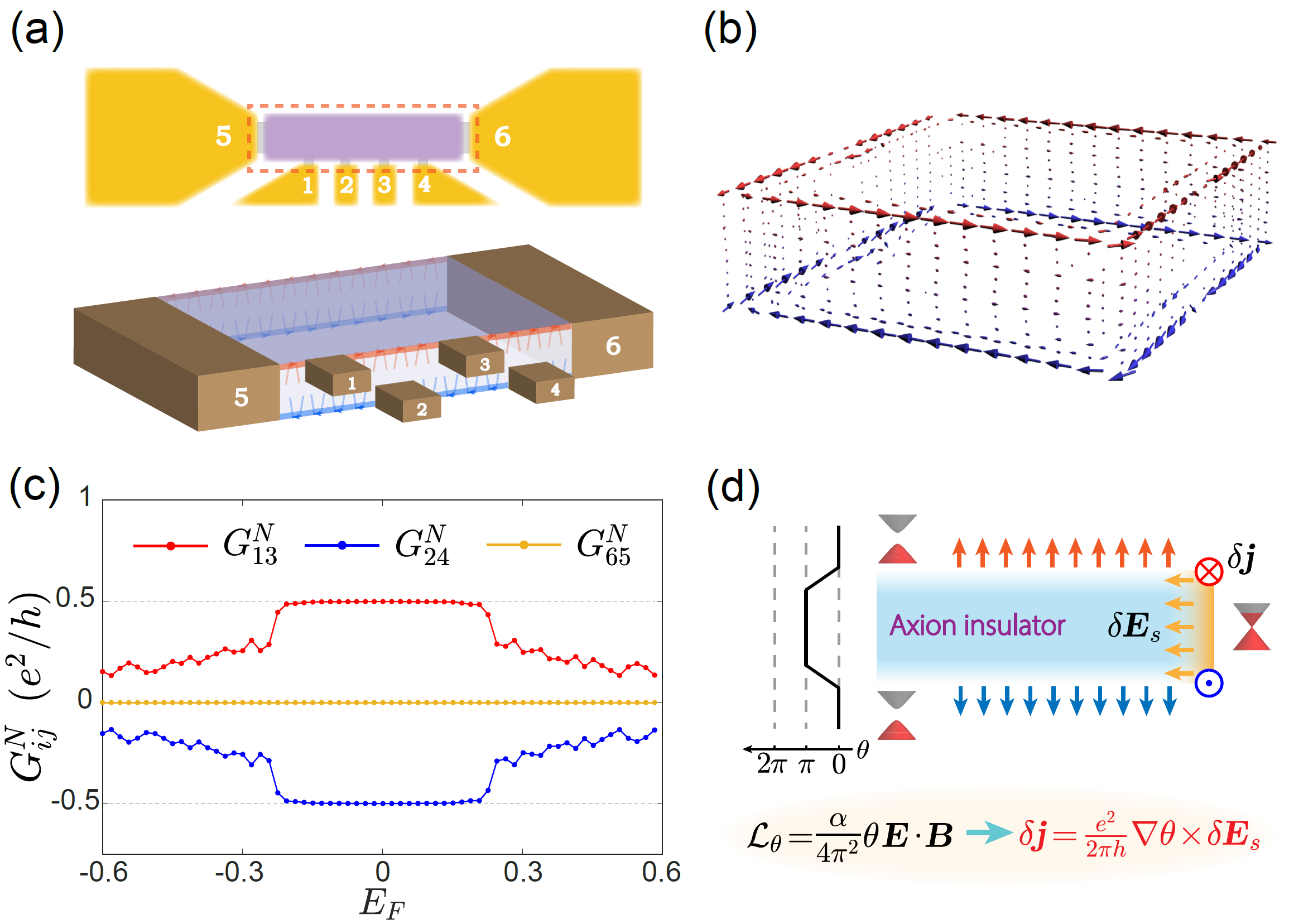}
\caption{(a) Schematic of the six terminal device. Leads 5 and 6 contact to the ends of the AI film. Terminals 1$\sim$4 are surface leads with 1 and 3 (2 and 4) contacted near the top (bottom) surface. (b) Local current distribution of the AI. (c) Nonreciprocal conductance $G^{N}_{ij}$ between lead $i$ and lead $j$ as a function of $E_{F}$. (d) Schematic illustration of the relationship between the quantized TME response and the half-quantized helical hinge currents. $\delta \bm{E}_{s}$ is the interface electric field and $\delta \bm{j}$ is the hinge current.}
\label{fig4}
\end{figure}

\emph{Experimental characterization of the half-quantized helical hinge currents.}--The half-quantized hinge current $\delta I_{GH}$ can be experimentally detected by a six-terminal device as illustrated in Fig.~\ref{fig4}(a). Leads $1$ and $3$ ($2$ and $4$) are contacted near the top (bottom) surface of the AI,
while leads $5$ and $6$ are contacted to the ends of the sample.
According to the Landauer-B\"{u}ttiker formula, the transmission coefficient between terminals $i$ and $j$ is $T_{ij}={\rm Tr}[{\mathbf \Gamma}_{i}{\mathbf G}^{r}{\mathbf \Gamma}_{j}{\mathbf G}^{a}]$ and the related differential conductance is $G_{ij}=e^{2}/hT_{ij}$ \cite{fisher_relation_1981,mackinnon_calculation_1985,metalidis_greens_2005,haug_hartmut_and_jauho_antti-pekka_and_others_quantum_2008}. The measurement method of $G_{ij}$ is given in \cite{Note1}. ${\mathbf \Gamma}_{i}$ is the linewidth function of lead $i$ and ${\mathbf G}^{r(a)}$ is the retarded (advanced) Green's function of the AI sample.

To demonstrate the existence of the helical hinge channels, we calculate the nonreciprocal conductance $G^{N}_{ij}=G_{ij}-G_{ji}$ between lead $i$ and lead $j$. As shown in Fig.~\ref{fig4}(c), in the AI state $G^{N}_{13}=e^{2}/2h$ and $G^{N}_{24}=-e^{2}/2h$ are half-quantized with opposite sign and $G^{N}_{65}=0$. This implies that there exist two counterpropagating half-quantized hinge channels. Moreover, the spatial distribution of the local current $J_{{\mathbf i}\to {\mathbf j}}(E)$ from site ${\mathbf i}$ to ${\mathbf j}$ \cite{jiang_numerical_2009} further reveals the helical signature of the transport hinge currents in the AI as shown in Fig.~\ref{fig4}(b).
Since the nonreciprocal conductance $G^{N}_{ij}$ counts the asymmetric part of $J_{{\mathbf i}\to {\mathbf j}}(E)$, we can deduce that the half-quantized conducting channel originates from the half-quantized chiral GH shift current $\delta I_{GH}$ on the hinge of the AI. The results are well consistent with those determined from GH shift (see Fig.~\ref{fig2}) as well as the current density distribution (see Fig.~\ref{fig3}). Therefore, these transport signals strongly confirm the reliability of the microscopic picture of half-quantized hinge currents proposed in Fig.~\ref{fig1}. Since the half-quantized helical hinge currents serve as the fingerprint of the AI, our proposals also promote the experimental identification of the AIs.

\emph{Discussion.}--The half-quantized helical hinge currents can be understood as a consequence of the quantized TME response in the AI \cite{qi_topological_2008}. As sketched in Fig.~\ref{fig4}(d), a shift of the Fermi energy on the surfaces of the AI induces an interface electric field $\delta\bm{E}_{s}=-\nabla \delta E_{F}/e$. According to the Maxwell equations modified by the axion term Lagrangian, $\delta \bm{j}=e^{2}/2\pi h\nabla \theta \times \delta\bm{E}_{s}$ \cite{qi_topological_2008,sitte_topological_2012} with the spatially varying axion angle $\theta$. The half-quantized hinge current $\delta I=\pm e/2h\delta E_{F}$ is obtained by integrating $\delta \bm{j}$ over the corner.

\emph{Conclusion.}--In conclusion, we found that the half-quantized helical hinge currents exist in AIs, and established their microscopic picture based on the GH shift currents. The half-quantization of the GH shift current has a topological origin and is robust to the variation of parameters. We numerically demonstrated that the half-quantized hinge channel is reflected by the half-quantized nonreciprocal conductances. Our studies deepen the perception of the boundary excitations of the AI and shed light on the detection of AIs through transport experiments.

\begin{acknowledgments}
We thank Qingfeng Sun, and Zhida Song for fruitful discussions. This work is financially supported by the National Basic Research Program of China (Grants No. 2019YFA0308403 and No. 2015CB921102), the National Natural Science Foundation of China (Grants No. 11974256), and the Strategic Priority Research Program of the Chinese Academy of Sciences (Grant No. XDB28000000). C.-Z. C. is also funded by the Natural Science Foundation of Jiangsu Province Grant No. BK20190813 and the Priority Academic Program Development (PAPD) of Jiangsu Higher Education Institutions.
\end{acknowledgments}
\bibliography{GH}

\end{document}

\title{Supplementary Materials for ``Half-Quantized Helical Hinge Currents in Axion Insulators''}
	
\author{Ming Gong}
\affiliation{International Center for Quantum Materials, School of Physics, Peking University, Beijing 100871, China}

\author{Haiwen Liu}
\affiliation{Center for Advanced Quantum Studies, Department of Physics, Beijing Normal University, Beijing 100875, China}

\author{Hua Jiang}
\email{jianghuaphy@suda.edu.cn}
\affiliation{School of Physical Science and Technology, Soochow University, Suzhou 215006, China.}
\affiliation{Institute for Advanced Study, Soochow University, Suzhou 215006, China.}
	
\author{Chui-Zhen Chen}
\email{czchen@suda.edu.cn}
\affiliation{School of Physical Science and Technology, Soochow University, Suzhou 215006, China.}
\affiliation{Institute for Advanced Study, Soochow University, Suzhou 215006, China.}
	
\author{X. C. Xie}
\email{xcxie@pku.edu.cn}
\affiliation{International Center for Quantum Materials, School of Physics, Peking University, Beijing 100871, China}
\affiliation{CAS Center for Excellence in Topological Quantum Computation,
		University of Chinese Academy of Sciences, Beijing 100190, China}

\maketitle
	
%\counterwithin{equation}{section} % reset the counting for the figure captions.
	
%\counterwithin{figure}{section} % reset the counting for the figure captions.
	
\tableofcontents
\section{Introduction for the Supplementary Materials}\label{sec:1Introduction for the Supplementary Materials}
In this supplementary material, we give detailed explanations of the models and methods used in the main text, and give detailed derivations of the analytical results. In \ref{sec:2Stationary phase method in calculating the GH shift}, we use the traditional stationary phase method to derive an analytical expression for the GH shift. In \ref{sec:3Probability flux method in calculating the GH shift}, we make necessary additions to the technical details of the probability flux method mentioned in the main text, and compare the evanescent wave and interference wave components of the GH shift as a function of $E$ at different incidence angles. In \ref{sec:4Anomalous velocity and band modification by the GH shift}, we derive the anomalous velocity induced by the GH shift and use a 2D lattice model to verify its modification to the band structure. In \ref{sec:5Derivation of the half-quantized GH shift current}, we use the anomalous velocities induced by the GH shift to derive a half-quantized GH shift current in the AI. In \ref{sec:6Power law decay of the interference wave part of the GH shift current}, we discuss the power low decay of the interference wave part of the GH shift current, which serves as a unique feature of the half-quantized hinge current. In \ref{sec:7Topological origin of the GH effect based on the adiabatic charge transport theory}, we use the adiabatic charge transport theory to give a topological understanding of the half-quantized GH shift current. In \ref{sec:8Derivation of the cross-section current density}, we derive the expression of the cross-section local current density. In \ref{sec:9Derivation of the differential conductance and the local current density}, we use the non-equilibrium Green's function method to derive the multi-terminal differential conductance and the local current density. In \ref{sec:10Chiral edge transport and its relation to half-quantized hinge channels in Chern insulators}, we additionally illustrate the relationship between chiral edge transport and half-quantized hinge channels in the CI, which is in stark contrast to the helical side surface transport in the AI phase. In \ref{sec:11Experimental setups to measure the nonreciprocal conductances}, we propose a feasible experimental setup to elucidate the principle in measuring nonreciprocal conductances. Finally, in \ref{sec:12Model parameters in numerics}, we list the model parameters used in the numerical calculations.
\section{Stationary phase method in calculating the GH shift}\label{sec:2Stationary phase method in calculating the GH shift}
We use the stationary phase method to derive the expression of the GH shift $\Delta_{GH}$ \cite{kurt_artmann_calculation_1948,beenakker_quantum_2009,jiang_topological_2015}. We consider the scattering problem described by the 2D Dirac Hamiltonian
\begin{equation}\label{eq:1}
\mathcal H(\bm{r})=\left\{
\begin{array}{ll}
\hbar v_{F}(-i\sigma_{x}\partial_{x}-i\sigma_{y}\partial_{y})-U&(x\leqslant 0)\\
\hbar v_{F}(-i\sigma_{x}\partial_{x}-i\sigma_{y}\partial_{y})+m\sigma_{z}&(x>0)
\end{array}\right.
\end{equation}
used in the main text. The incident plane wave and the reflected plane wave are $\psi_{\rm{in}}(\bm{r})=e^{ik_{x}x+ik_{y}y}[e^{-i\frac{\alpha}{2}},e^{i\frac{\alpha}{2}}]^{T}/\sqrt{2}$ and $\psi_{\rm{re}}(\bm{r})=e^{-ik_{x}x+ik_{y}y}[e^{-i\frac{\pi-\alpha}{2}},e^{i\frac{\pi-\alpha}{2}}]^{T}/\sqrt{2}$. The scattering problem is solved by matching the boundary conditions of incident/reflected wave $\psi(\bm{r})=\psi_{\rm{in}}+r\psi_{\rm{re}}$ ($x\leqslant 0$) and the penetrated evanescent wave $\psi_{{\rm eva}}(\bm{r})=e^{-\kappa x+ik_{y}y}\psi(0,0)$ ($x>0$), and obtain the reflection coefficient $r=e^{i\phi_{r}}$. We have
\begin{eqnarray}
r=e^{i\phi_{r}}=\frac{me^{-i\frac{\alpha}{2}}+i(\kappa-k_{y})e^{i\frac{\alpha}{2}}-Ee^{-i\frac{\alpha}{2}}} {ime^{i\frac{\alpha}{2}}+(\kappa-k_{y})e^{-i\frac{\alpha}{2}}-iEe^{i\frac{\alpha}{2}}},
\end{eqnarray}
where $\kappa=\sqrt{k_{y}^{2}+m^{2}-E^{2}}$, $E=\sqrt{k_{x}^{2}+k_{y}^{2}}-U$, and $\alpha={\rm arctan}\frac{k_{y}}{k_{x}}$. Therefore, $\phi_{r}$ can be viewed as a function of $k_{y}$ or $\alpha$ for fixed $E$, $U$, and $m$. 
	
The incident and reflected Gaussian wave packets are constructed as
\begin{eqnarray}
\psi^{{\rm in}}_{g}(\bm{r})&&=\int dk_{y}\frac{1}{\sqrt{2\pi}\Delta_{k_{y}}}{\rm exp}[-\frac{(k_{y}-\bar{k_{y}})^{2}}{2\Delta_{k_{y}}^2}]e^{ik_{x}x+ik_{y}y}\frac{1}{\sqrt{2}}[e^{-i\frac{\alpha}{2}},e^{i\frac{\alpha}{2}}]^{T},\\
\psi^{{\rm re}}_{g}(\bm{r})&&=\int dk_{y}\frac{1}{\sqrt{2\pi}\Delta_{k_{y}}}{\rm exp}[-\frac{(k_{y}-\bar{k_{y}})^{2}}{2\Delta_{k_{y}}^2}]e^{-ik_{x}x+ik_{y}y+i\phi_{r}}\frac{1}{\sqrt{2}}[e^{-i\frac{\pi-\alpha}{2}},e^{i\frac{\pi-\alpha}{2}}]^{T}.
\end{eqnarray}
Note that the integration is only performed for $k_{y}$ because the eigen equation $\mathcal H\psi^{{\rm in/re}}_{g}=E\psi^{{\rm in/re}}_{g}$ constraints the number of free variables through $k_{x}=\sqrt{(E+U)^{2}-k_{y}^{2}}$ and $\alpha={\rm arctan}\frac{k_{y}}{k_{x}}$ are functions of $k_{y}$. Expand $\alpha(k_{y})$ and $\phi_{r}(k_{y})$ to the first order of $(k_{y}-\bar k_{y})$ around $\bar k_{y}$ as $\alpha=\bar\alpha +(k_{y}-\bar k_{y})\partial\alpha/\partial k_{y}+O((k_{y}-\bar k_{y})^{2})$ and $\phi_{r}=\bar\phi_{r} +(k_{y}-\bar k_{y})\partial\bar\phi_{r}/\partial k_{y}+O((k_{y}-\bar k_{y})^{2})$. Then substitute them into Eq. (S3) and Eq. (S4) we have (accurate to the first order of $k_{y}$)
\begin{eqnarray}
\psi^{{\rm in}}_{g,\pm}(\bm{r})&&\propto e^{-(y\mp\frac12 \frac{\partial \bar\alpha}{\partial k_{y}})^2\Delta_{k_{y}}^2/2}\\
\psi^{{\rm re}}_{g,\pm}(\bm{r})&&\propto e^{-(y+\frac{\partial \bar\phi_{r}}{\partial k_{y}}\pm\frac12 \frac{\partial \bar\alpha}{\partial k_{y}})^2\Delta_{k_{y}}^2/2}
\end{eqnarray}
for spin up (+) and spin down ($-$) components. From now on we replace $\bar\phi_{r}$, $\bar k_{y}$, and $\bar\alpha$ with $\phi_{r}$, $k_{y}$, and $\alpha$. It is clear from Eq. (S5) and Eq. (S6) that the GH shift is the displacement of the wave packet center. GH shifts for spin up and spin down components are $\Delta_{GH}^{+}=-\partial\phi_{r}/\partial k_{y}-\partial\alpha/\partial k_{y}$ and $\Delta_{GH}^{-}=-\partial\phi_{r}/\partial k_{y}+\partial\alpha/\partial k_{y}$. The total GH shift is the spin averaged wave packet center displacement $\Delta_{GH}= (\Delta^{+}_{GH}+\Delta^{-}_{GH})/2=-\frac{\partial \phi_{r}}{\partial k_{y}}$. Substitute it into Eq. (S2) we have
\begin{eqnarray}
\Delta_{GH}&&=-\frac{\partial \phi_{r}}{\partial k_{y}}=2\frac{m\kappa(m-E)+\kappa^2(k_{y}-\kappa)+(E-m)(k_{y}-\kappa)k_{x}{\rm cos}\alpha}{\kappa k_{x} [(E-m)^2+(\kappa-k_{y})^2+2(\kappa-k_{y})(E-m){\rm sin}\alpha]}.
\end{eqnarray}
\section{Probability flux method in calculating the GH shift}\label{sec:3Probability flux method in calculating the GH shift}
The probability flux method is originally named ``the energy flux method" in studying the GH shift of light beams based on the energy conservation condition \cite{renard_total_1964,yasumoto_new_1998}. Here, we treat the quantum wave system and replace the energy flux with the probability flux.    
\begin{figure}[t]
	\includegraphics[width=1\textwidth]{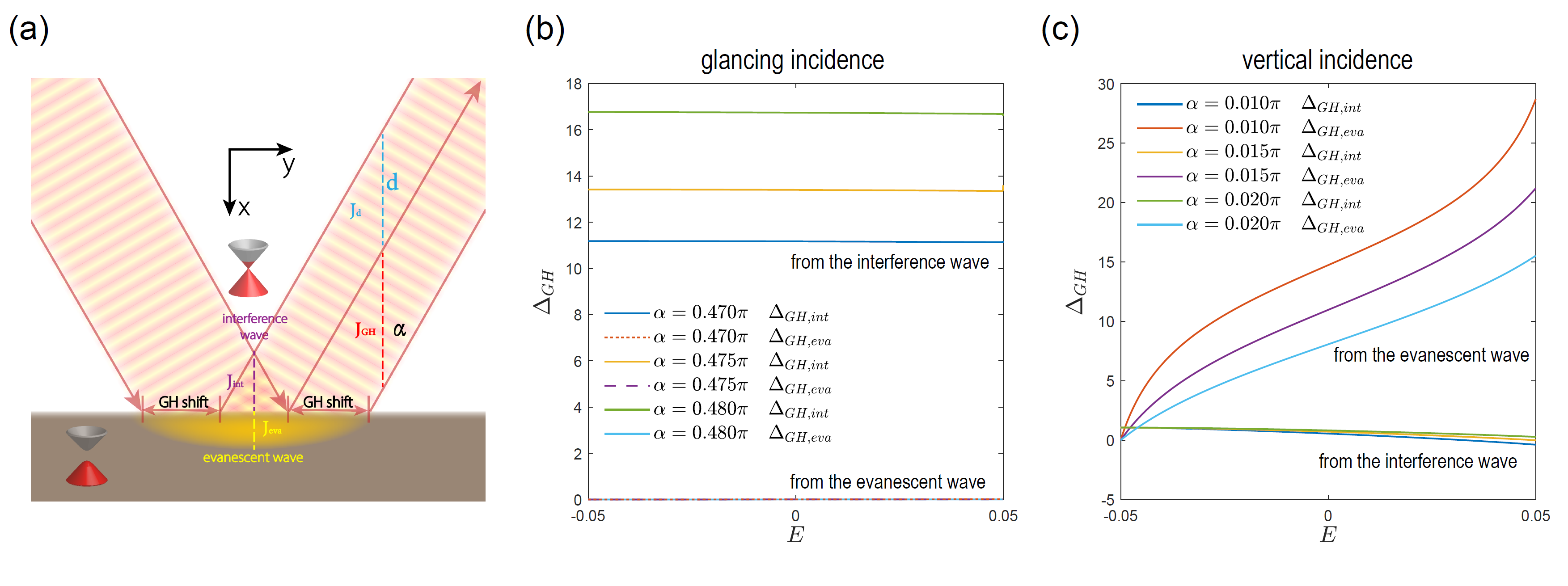}
	\caption{(a) Sketch of the scattering process where a massless Dirac electron bouncing off a massive barrier, where $J_{\rm{int}}$ ($J_{\rm{eva}}$) represents the flux carried by the interference wave (evanescent wave), $J_{\rm{GH}}$ represents the flux induced by GH shift, and $J_{\rm{d}}$ represents the flux  through the cross-section colored blue. (b) $\Delta_{GH,int}$ and $\Delta_{GH,eva}$ as a function of $E$ for near vertical incidence. (c) $\Delta_{GH,int}$ and $\Delta_{GH,eva}$ as a function of $E$ for near glancing incidence.}  
	\label{s1}
\end{figure}
Suppose the probability density of the incident/reflected beam is normalized to one, i. e. $\psi_{\rm in}^{\dagger}({\bm r})\psi_{\rm in}({\bm r})=\psi_{\rm re}^{\dagger}({\bm r})\psi_{\rm re}({\bm r})=1$, then 
\begin{eqnarray}
J_{\rm{d}}&&=v_{F}d\rm{sin}\alpha,\\
J_{\rm{GH}}&&=v_{F}\Delta_{GH}\rm{cos}\alpha,\\
J_{\rm{eva}}&&=\int_{0}^{\infty}dx\psi_{\rm{eva}}^{\dagger}v_{F}\sigma_{y}\psi_{\rm{eva}}=v_{F}({\rm sin}\alpha+{\rm cos}\phi_{r})/\kappa,
\end{eqnarray}
where $\psi_{{\rm eva}}(\bm{r})=e^{-\kappa x+ik_{y}y}\psi_{\rm int}(0,0)$ for $x> 0$ and $\psi_{\rm int}(\bm{r})=\psi_{\rm{in}}+r\psi_{\rm{re}}$ for $x\leqslant 0$. [See Fig.~\ref{s1}(a)]. The flux conservation condition implies that $J_{\rm{int}}+J_{\rm{eva}}=J_{\rm{GH}}+J_{\rm{d}}$. Following the standard flux method analysis, 
\begin{eqnarray}
\nonumber
J_{\rm{int}}(d)&&= \int_{-d/2}^{0}dx\psi_{\rm int}^{\dagger}v_{F}\sigma_{y}\psi_{\rm int}\\ \nonumber
&&=\int_{-d/2}^{0}dxv_{F}[2{\rm cos}(\phi_{r}-2k_{x}x)+2{\rm sin}\alpha]\\
&&=v_{F}d{\rm sin}\alpha+v_{F}\frac{{\rm sin}(\phi_{r}+k_{x}d)}{k_{x}}-v_{F}\frac{{\rm sin}\phi_{r}}{k_{x}}.
\end{eqnarray} 
Therefore, the GH shift as a function of $d$ can be obtained as
\begin{eqnarray}
\Delta_{GH}(d)=\frac{{\rm sin}\alpha+{\rm cos}\phi_{r}}{\kappa {\rm cos}\alpha}-\frac{{\rm sin}\phi_{r}}{k_{x}{\rm cos}\alpha}+\frac{{\rm sin}(\phi_{r}+k_{x}d)}{k_{x}{\rm cos}\alpha}.
\end{eqnarray}
The $d$ dependence of $\Delta_{GH}(d)$ only comes from the last term in Eq . (S12), which oscillates with $d$ and can be averaged out. Then we have
\begin{eqnarray}
\Delta_{GH}=\langle\Delta_{GH}(d)\rangle_{d}=\frac{{\rm sin}\alpha+{\rm cos}\phi_{r}}{\kappa {\rm cos}\alpha}-\frac{{\rm sin}\phi_{r}}{k_{x}{\rm cos}\alpha}.
\end{eqnarray}
It can be verified that Eq. (S7) and Eq. (S13) are equivalent.
	
It is clear that the first term in Eq. (S13) comes from the evanescent wave and the second term comes from the interference wave. We plot $\Delta_{GH,eva}$ and $\Delta_{GH,int}$ as a function of $E$ for different incident angles $\alpha$ in Fig.~\ref{s1}(b) and (c). We can see that for small incident angles (near vertical incidence), the total GH shift is mostly contributed from the evanescent wave while for large incident angles (near glancing incidence) the total GH shift is mostly contributed from the interference wave.
\section{Anomalous velocity and band modification by the GH shift}\label{sec:4Anomalous velocity and band modification by the GH shift}
\begin{figure}[htbp]
	\includegraphics[width=1\textwidth]{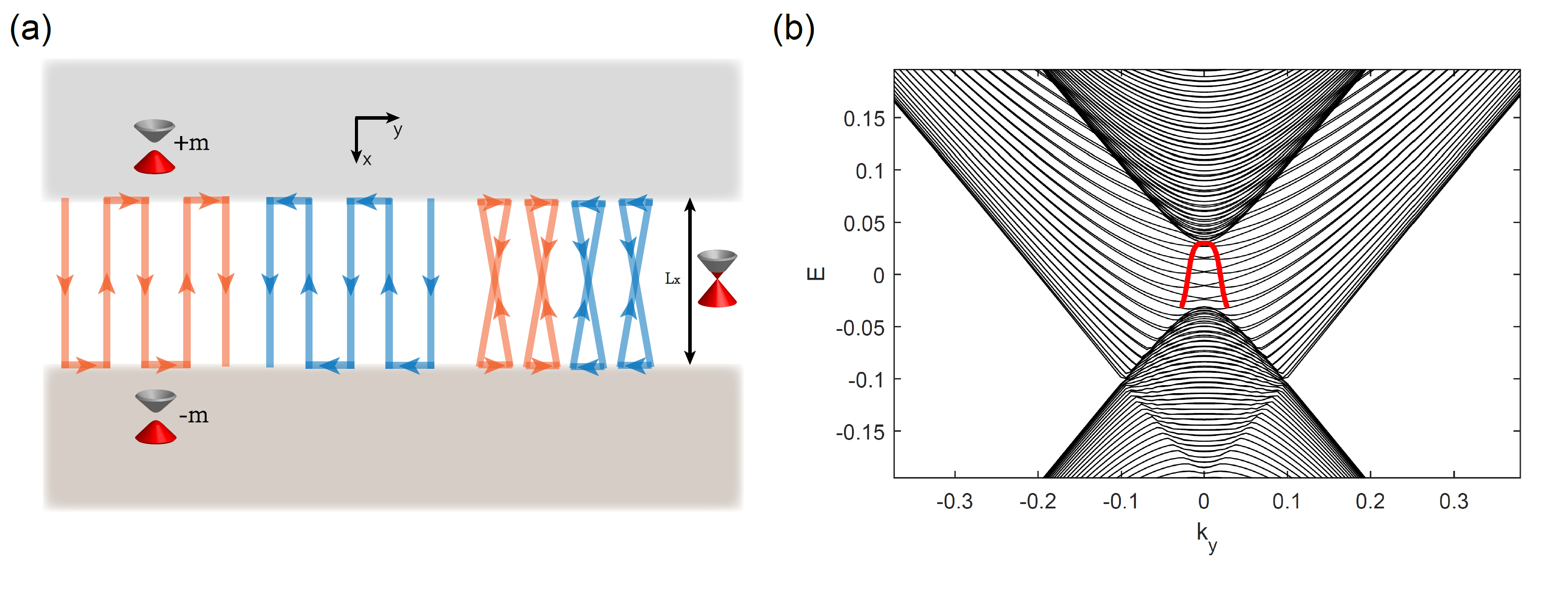}
	\caption{(a) Sketch of the process where massless Dirac electron bounces back and forth between massive barriers with opposite mass. Trajectories colored blue and red represent electrons from difference valleys at $k_{y}=0$ [$(0,0)$ and $(0,\pi)$]. (b) Band structure modification by the GH shift induced anomalous velocity. The band is numerically obtained from a tight binding model ($\mu=0.2$, $L_{x}=600$, and $m=0.03$). The red curve represents the predicted band minimum calculated from the GH shift, which suggests the existence of the zero-velocity state at non-zero $k_{y}$.}  
	\label{s2}
\end{figure}
Before deriving the half-quantized GH shift current in AIs and CIs, we investigate the anomalous velocity induced by the GH shift and its modification to the band structure. Consider a massless Dirac electron bounces back and forth between massive barriers with opposite mass as shown in Fig.~\ref{s2}(a). The average time interval between two consecutive bounces is $\Delta \tau=\frac{L_{x}}{v_{F}{\rm cos}\alpha}$. The anomalous velocity induced by the GH shift along the $y$ direction is
\begin{eqnarray}
v_{GH}=\frac{\Delta_{GH}}{\Delta \tau}=\frac{\Delta_{GH}v_{F}{\rm cos}\alpha}{L_{x}},\quad (-\pi/2<\alpha<\pi/2)
\end{eqnarray}
For electrons with incident angle $\alpha$, the drift velocity (without the GH shift induced anomalous velocity) is $v_{0}=v_{F}{\rm sin}\alpha$.
The total velocity vanishes at the band minimum $v_{0}+v_{GH}=0$.
We then have the condition for the band minimum
\begin{eqnarray}
-\frac{\Delta^{y}}{L_{x}{\rm tan}\alpha}=1.
\end{eqnarray}
The band minimum indicates the existence of the ``8'' shape trajectories as sketched in Fig.~\ref{s2}(a).
	
We use the 2D tight binding Hamiltonian \cite{marchand_lattice_2012}
\begin{eqnarray}
H_{2D}=\sum_{\mathbf{i}}\Big[\frac{i \hbar v_{F}}{2 a}(c_{\mathbf{i}}^{\dagger} \sigma_{y} c_{\mathbf{i}+\delta \hat{\mathbf{x}}}-c_{\mathbf{i}}^{\dagger} \sigma_{x} c_{\mathbf{i}+\delta \hat{\mathbf{y}}})-\frac{\mu}{2}c_{\mathbf{i}}^{\dagger}\sigma_{0} c_{\mathbf{i}}+mc_{\mathbf{i}}^{\dagger}\sigma_{z} c_{\mathbf{i}}\Big]+\mathrm{H.c.}
\end{eqnarray}
to numerically investigate the band structure changes. We take $y$ direction to be infinite and in the $x$ direction the massless electrons with $m=0,\mu\neq0$ sandwiched between two barriers with opposite $m$ and $\mu=0$. The lattice Hamiltonian described by Eq. (S16) contains 4 valleys in total at $(0,0)$, $(0,\pi)$, $(\pi,0)$, and $(\pi,\pi)$. We investigate the two valleys at $k_{y}=0$ [$(0,0)$ and $(0,\pi)$]. Electrons from both valleys accumulate anomalous velocity when they bounce off the massive barrier, but with opposite direction as sketched in Fig.~\ref{s2}(a). From Fig.~\ref{s2}(b) we can see that band minimums for electrons from different valleys splits, which coincide with the analytical prediction based on Eq. (S15).
\section{Derivation of the half-quantized GH shift current}\label{sec:5Derivation of the half-quantized GH shift current}
\begin{figure}[htbp]
	\includegraphics[width=1\textwidth]{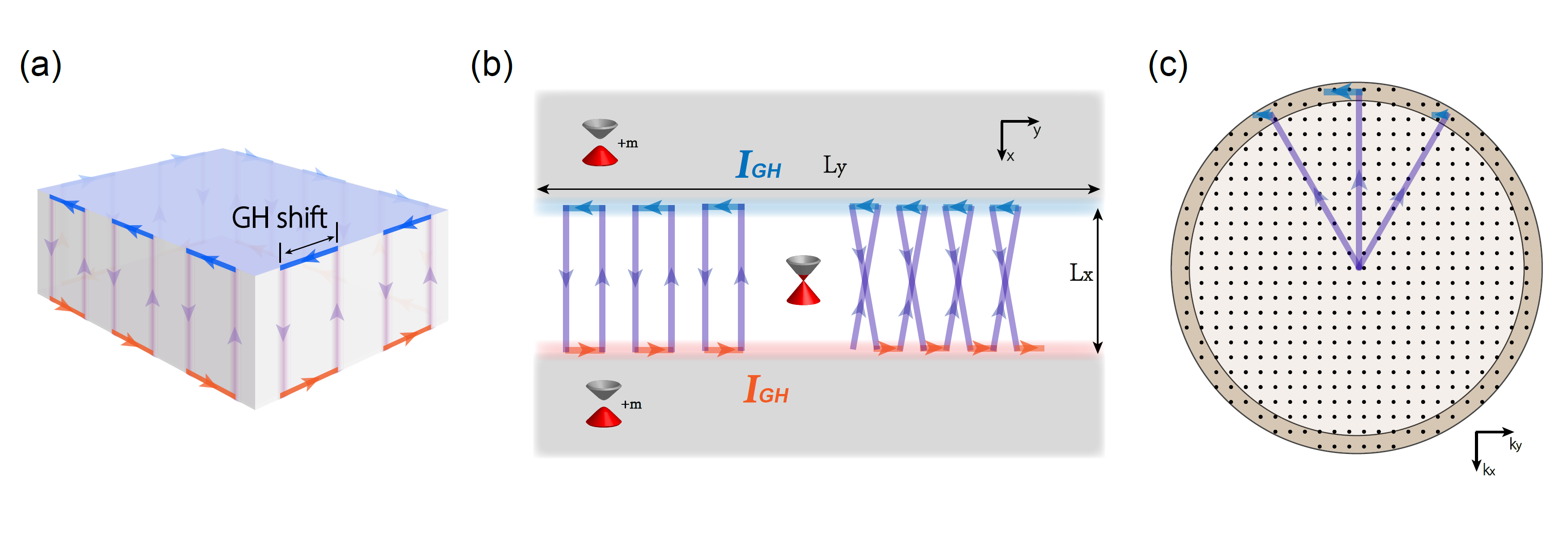}
	\caption{(a) Sketch of the GH shift current induced chiral hinge current in the AI. (b) Expanded view of the top-side-bottom surfaces of the AI. Massless Dirac electrons bounce back and forth between massive barriers with the same mass $m$, accumulate an anomalous shift current due to the chiral GH shift. $L_{x}$ and $L_{y}$ represent the width and length of the side surface for box normalization (in calculations we take the $y$ direction to be periodic). (c) Box normalization to count the GH shift contributions from electrons near the Fermi surface.}  
	\label{s3}
\end{figure}
We derive the half-quantized hinge current induced by the GH shift. Suppose the width between the barriers is $L_{x}$, the average time between two consecutive bounces off one of the barriers is $\Delta \tau =2L_{x}/v_{F}|\rm{cos}\alpha|$ for $-\pi<\alpha \leqslant\pi$. (Here, we consider the time interval between two consecutive bounces of electrons off the same barrier, thus the distance traveled in the $x$ direction is $2L_{x}$. Contributions from electrons with incident angle $\alpha$ and $\pi-\alpha$ are considered to be equivalent between successive bounces, i. e. $\Delta_{GH}(\alpha)=\Delta_{GH}(\pi-\alpha)$, since the electron with incident angle $\alpha$ will be alternated to $\pi-\alpha$ after one reflection. Therefore, we take $|\rm{cos}\alpha|$ to describe both cases.) The lateral GH shift $\Delta_{GH}$ induces an anomalous velocity of electrons near the barrier as $v_{GH}=\Delta_{GH}/\Delta\tau=\Delta_{GH}v_{F}|{\rm cos}\alpha|/2L_{x}(-\pi < \alpha \leqslant \pi)$. The total GH shift current is obtained by counting the contributions of all filled states as
\begin{eqnarray}
I_{GH}&&=\sum_{\rm filled} \frac{ev_{GH}}{L_{y}}=\sum_{\rm filled} \frac{e\Delta_{GH}v_{F}|{\rm cos}\alpha|}{2L_{x}L_{y}}.
\end{eqnarray}
Here, we take the box normalization for electrons on the side surface with length $L_{y}$ and width $L_{x}$, as depicted in Fig.~\ref{s3}(b). According to the box normalization [Fig.~\ref{s3}(c)], $\frac{1}{L_{x}L_{y}}\sum_{{\bm k}}=\frac{1}{(2\pi)^{2}}\int d{\bm k}$.
\begin{eqnarray}
I_{GH}&&=\int_{\rm filled} \frac{L_{x}L_{y}}{(2\pi)^{2}}KdKd\alpha\frac{e\Delta_{GH}(\alpha)v_{F}|{\rm cos}\alpha|}{2L_{x}L_{y}}\\
&&=\int_{-\infty}^{E_{F}}\frac{dE}{\hbar v_{F}}\int_{-\pi}^{\pi}\frac{Kd\alpha|{\rm cos}\alpha|}{(2\pi)^{2}}\frac{e\Delta_{GH}(\alpha)v_{F}}{2}\\
\nonumber
&&=\int_{-\infty}^{E_{F}}dE\cdot 2\int_{-K}^{K}\frac{dk_{y}}{2\pi}\frac{e\Delta_{GH}(k_{y},E)}{2\cdot2\pi\hbar}\\
&&=\int_{-\infty}^{E_{F}}dE\frac{e}{h}\int_{-K}^{K}\frac{dk_{y}}{2\pi}\Delta_{GH}(k_{y},E).
\end{eqnarray}
From Eq. (S18) to Eq. (S19), we used the fact that $\int_{-\pi}^{\pi}\Delta_{GH}(\alpha)|{\rm cos}\alpha|d\alpha=2\int_{-\pi/2}^{\pi/2}\Delta_{GH}(\alpha){\rm cos}\alpha d\alpha=2\int_{-\pi/2}^{\pi/2}\Delta_{GH}(\alpha)d{\rm sin}\alpha$ since we take $\Delta_{GH}(\alpha)=\Delta_{GH}(\pi-\alpha)$. $K=(E+U)/\hbar v_{F}$ is the Fermi wave vector at energy $E$. 
	
Some points should be noted for Eq. (S19) and Eq. (S20). In Eq. (S19), the integrand $v_{F}\Delta_{GH}{\rm cos}\alpha$ is exactly the GH shift induced probability flux $J_{GH}$ in Eq. (S9). From the flux conservation perspective $J_{GH}=J_{eva}+J_{int}-J_{d}$, the GH shift current have both contributions from the evanescent wave and the interference wave, indicating that our derivations of the GH shift as well as the GH shift current are equivalent. In Eq. (S20), the integration over $E$ is not necessarily performed from $-\infty$ to $E_{F}$, since we only derived the $\Delta_{GH}$ for the total reflection case within the gap. The only thing we are interested is the differential GH shift current with respect to $E_{F}$, which contributes to the transport current.  From Eq. (S7) $\Delta_{GH}=-\partial\phi_{r}/\partial k_{y}$ and we have the differential form of $I_{GH}$
\begin{eqnarray}
\delta I_{GH}=\delta E_{F}\frac{e}{h}\int_{-K}^{K}\frac{dk_{y}}{2\pi}\Delta_{GH}(k_{y},E)=\delta E_{F}\frac{e}{h}[\phi_{r}(-\pi/2)-\phi_{r}(\pi/2)]/2\pi
\end{eqnarray}
for $E_{F}$ lying in the gap of the top and bottom surfaces. From the plot in Fig.~2(d) in the main text, we immediately obtain the half-quantized GH shift current $\delta I_{GH}=\frac{e}{2h}\delta E_{F}$.

\section{Power law decay of the interference wave part of the GH shift current}\label{sec:6Power law decay of the interference wave part of the GH shift current}	
In the maintext and \ref{sec:3Probability flux method in calculating the GH shift} we discussed the decomposition of the GH shift current according to the contributions from the evanescent wave part and the interference wave part. In this section, we emphasize that the	interference wave induced GH shift current component decays from the edge following the power law, which is in stark contrast to the current carried by the topological edge or hinge state that decays exponentially. To see clearly, we now turn to the integrand in Eq. (S11). The $y$-component of the current as a function of $x$ carried by the interference wave is $j_{int}(x)=\psi_{\rm int}^{\dagger}v_{F}\sigma_{y}\psi_{\rm int}(x)=2v_{F}[{\rm cos}(\phi_{r}-2k_{x}x)+{\rm sin}\alpha]$, where $k_{x}=K\cos \alpha$ and $K=(E_{F}+U)/\hbar v_{F}$ is the Fermi wave vector. The total contribution of the current from all the $\alpha$ can be written as
\begin{eqnarray}
\nonumber
\mathcal{J}_{int}(x)&&=\int_{-\frac{\pi}{2}}^{\frac{\pi}{2}} j_{int}(x)d\alpha\propto \int_{-\frac{\pi}{2}}^{\frac{\pi}{2}} [{\rm cos}(\phi_{r}-2k_{x}x)+{\rm sin}\alpha]d\alpha\\
&&= \int_{-\frac{\pi}{2}}^{\frac{\pi}{2}}{\rm cos}(\phi_{r}-2k_{x}x)d\alpha.
\end{eqnarray} 
From Eq. (S2), if $m\to \infty$, then $\phi_{r}\to 0$ for most of the $\alpha$. We omit the $\phi_{r}$ for simplicity to study the asymptotic behavior of Eq. (S22). Then we have
\begin{eqnarray}
\mathcal{J}_{int}(x)\propto \int_{-\frac{\pi}{2}}^{\frac{\pi}{2}} {\rm cos}(2k_{x}x)d\alpha=\frac{1}{2}\int_{0}^{2\pi}{\rm cos}(2K{\rm cos}\alpha x)d\alpha=\frac{1}{4\pi}{\rm J}_{0}(2Kx).
\end{eqnarray}
Here, ${\rm J}_{0}(x)$ is the 0$_{\rm th}$ Bessel function. According to the asymptotic formula of the Bessel function
\begin{eqnarray}
\nonumber
\mathrm{J}_{n}(x) &&=\frac{1}{\pi}{\rm Re}\left[e^{-i \frac{n \pi}{2}} \int_{-\frac{\pi}{2}}^{\frac{\pi}{2}} e^{i x \cos \varphi} \cos n \varphi \mathrm{d} \varphi\right] \\
&&=\sqrt{\frac{2}{\pi x}} \cos \left(x-\frac{n \pi}{2}-\frac{\pi}{4}\right)+O(x^{-\frac{3}{2}}),
\end{eqnarray}
one can see that when $x\to \infty$
\begin{eqnarray}
\mathcal{J}_{int}(x)\propto {\rm J}_{0}(2Kx)\to \sqrt{\frac{1}{\pi Kx}} \cos \left(2Kx-\frac{\pi}{4}\right).
\end{eqnarray}
The result clearly shows that the interference wave part of the GH shift current maximizes at the boundary and decays to zero in a power law $x^{-\frac{1}{2}}$ when moving away from the boundary with the oscillation length as $1/2K$.
Note that the result in \cite{zou_half-quantized_2022} is similar to ours, where they conclude that the edge current decays to zero in a power law $x^{-\frac{3}{2}}$. The difference is that they counted the contributions from all states below the Fermi surface, while our result only focus on the states near the Fermi surface. Nevertheless, the results are consistent becasue the derivation of their result with respect to $K$ 
\begin{eqnarray}
\frac{d}{dK}\left[x^{-\frac{3}{2}} \cos \left(2Kx-\frac{3\pi}{4}\right)\right]\to x^{-\frac{1}{2}} \cos \left(2Kx-\frac{\pi}{4}\right)
\end{eqnarray}
when $x\to \infty$ is the same as ours.

Here we emphasize that the chiral current carried by the interference wave on the metallic side surface depends on the gapped, time-reversal symmetry breaking top/bottom surface. The power low decay of the current induced by the interference wave indicates that it cannot be generated by any kind of topologically protected edge or hinge state which decays exponentially. It also indicates that the half-quantized hinge current cannot exist by itself and should be combined with another one to form quantized side surface transport in the AI or the CI. These features are unique for the half-quantized hinge current.
\section{Topological origin of the GH effect based on the adiabatic charge transport theory}\label{sec:7Topological origin of the GH effect based on the adiabatic charge transport theory}
\begin{figure}[htbp]
	\includegraphics[width=1\textwidth]{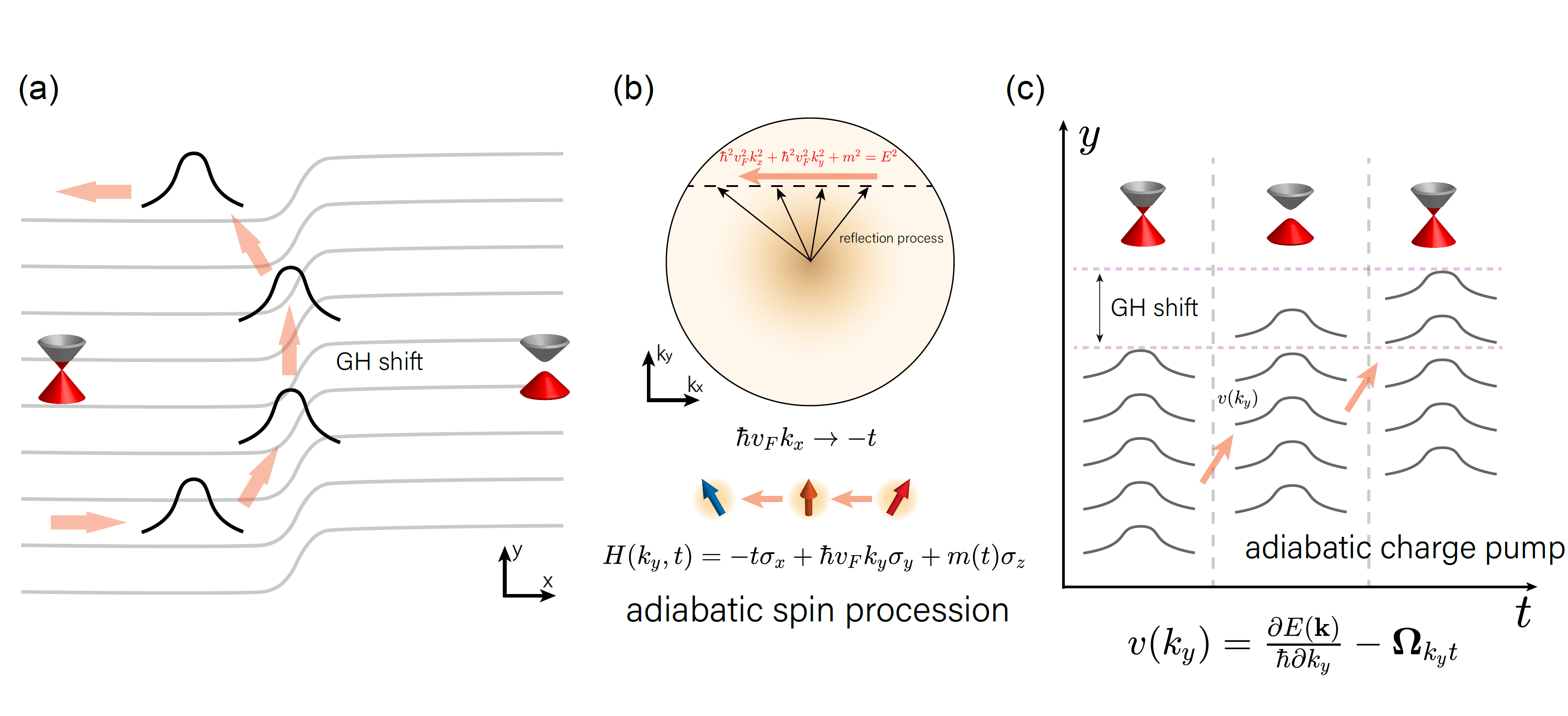}
	\caption{(a) Sketch of the process where a wave packet of a massless Dirac electron bounces off a massive barrier, undergoing a lateral GH shift in the $y$ direction. (b) Sketch of the reflection process of the wave packet in the momentum space where $k_{y}$ is unchanged due to the translation symmetry in the $y$ direction. During the reflection the energy $E$ is conserved, therefore $k_{x}$ and the mass $m$ of the local Hamiltonian is varying under the constraint $\hbar^{2}v_{F}^{2}k_{x}^{2}+\hbar^{2}v_{F}^{2}k_{y}^{2}+m^{2}=E^{2}$. Such a reflection process can also be understood as the 1D charge pumping problem in the $y$ direction when we take $\hbar v_{F}k_{x}=-t$ as the virtual time. The effective time-dependent Hamiltonian can be viewed as a Zeeman type $H={\bm B}\cdot{\bm \sigma}$ with ${\bm B}=(-t,\hbar v_{F}k_{y},m(t))$ and the reflection process is simplified to an adiabatic spin procession for fixed $k_{y}$ and $E$. (c) Sketch of the charge pumping process in the $y$ direction. The non-trivial Berry curvature $\Omega_{k_{y}t}$ induces a anomalous velocity $v(k_{y})$ in the $y$ direction. After the reflection, the total contribution from the adiabatic current gives rise to the GH shift.}  
	\label{s4}
\end{figure}
In this section, we give a semiclassical understanding of the topological origin of the GH effect based on the adiabatic charge transport theory \cite{thouless_quantized_1982,thouless_localisation_1981,thouless_quantization_1983,xiao_berry_2010}. Instead of Eq. (S1) that contains a sharp boundary between the massless and massive Dirac electron, we use Hamiltonian
\begin{equation}
\mathcal H(\bm{r})=	\hbar v_{F}(-i\sigma_{x}\partial_{x}-i\sigma_{y}\partial_{y})+m(x)\sigma_{z},\\
\end{equation}
where $m(x)$ is a smooth function connecting the gapless and the gapped regions with $m(x)\to m$ for $x\gg 0$ and $m(x)\to 0$ for $x\ll 0$. Here we drop the $U$ term in Eq. (S1) without affecting the conclusion. When the domain wall is large enough, the motion of the wave packet can be viewed semiclassically with specific momentum ${\bm k}$ and position ${\bm r}$ [see Fig.~\ref{s4}(a)]. The local Hamiltonian reads
\begin{equation}
H(\bm{k})=	\hbar v_{F}(k_{x}\sigma_{x}+k_{y}\sigma_{y})+m(x)\sigma_{z}.
\end{equation}
During the scattering process, the energy $E$ and momentum in the $y$ direction $k_{y}$ are unchanged. Therefore, the relation holds
\begin{eqnarray}
\hbar^{2}v_{F}^{2}k_{x}^{2}+\hbar^{2}v_{F}^{2}k_{y}^{2}+m^{2}=E^{2}.
\end{eqnarray} 
	
To map the scattering problem into a charge transfer problem, we take $\hbar v_{F} k_{x}= -t$ as the virtual time. The local Hamiltonian in Eq. (S28) becomes time-dependent
\begin{eqnarray}
H(k_{y},t)=-t\sigma_{x}+\hbar v_{F}k_{y}\sigma_{y}+m(t)\sigma_{z},
\end{eqnarray}
which describes the Zeeman coupling of a Pauli spinor to a time-dependent magnetic field ${\bm B}=(-t,\hbar v_{F}k_{y},m(t))$ [see Fig.~\ref{s4}(b)]. Then, the reflection process is reduced to an adiabatic spin procession. We denote the instantaneous eigen states of Eq. (S30) as $|u_{\pm}(k_{y},t)\rangle$, where $\pm$ denotes the spin up and spin down components of the spinor. It is easy to solve the instantaneous eigenequation $H(k_{y},t)|u_{\pm}(k_{y},t)\rangle=E_{\pm}|u_{\pm}(k_{y},t)\rangle$ and obtain $|u_{\pm}(k_{y},t)\rangle=[E\pm m,-t+i\hbar v_{F} k_{y}]^{T}/\sqrt{2E(E\pm m)}$. Following the analysis in \cite{xiao_berry_2010}, apart from an unimportant overall phase factor up to the first order in the rate of the change in the Hamiltonian, the wave function is given by 
\begin{eqnarray}
|u_{\pm}(k_{y},t)\rangle-i\hbar\sum_{n'\neq n}\frac{|u_{n'}(k_{y},t)\rangle \langle u_{n'}(k_{y},t)|\partial u_{n}(k_{y},t)/\partial t\rangle}{E_{n}-E_{n'}},
\end{eqnarray}
where $n(n')=\pm$ represents the spin up or spin down components, and can further be viewed as the band index in a 1D (in the $y$ direction) two-band model Eq. (S30). The average velocity for a given $k_{y}$ is found to the first order
\begin{eqnarray}
v_{n}(k_{y})&&=\partial E_{n}(k_{y})/\hbar\partial k_{y}-i\sum_{n'\neq n}\left\{ \frac{\langle u_{n}|\partial H/\partial k_{y}|u_{n'}\rangle \langle u_{n'}|\partial u_{n}/\partial t\rangle}{E_{n}-E_{n'}}-c.c. \right\}\\
&&=\partial E_{n}(k_{y})/\hbar\partial k_{y}-i\left[ \left \langle \frac{\partial u_{n}}{\partial k_{y}}\bigg| \frac{\partial u_{n}}{\partial t} \right\rangle- \left \langle \frac{\partial u_{n}}{\partial t}\bigg| \frac{\partial u_{n}}{\partial k_{y}} \right\rangle\right]\\
&&=\partial E_{n}(k_{y})/\hbar\partial k_{y}-\Omega^{n}_{k_{y}t}.
\end{eqnarray}
Here, we used the relation $\left\langle u_{n}|\partial H / \partial k_{y}| u_{n^{\prime}}\right\rangle=\left(E_{n}-E_{n^{\prime}}\right)\left\langle\partial u_{n} / \partial k_{y}|u_{n^{\prime}}\right\rangle$ and the identity $\sum_{n^{\prime}}\left|u_{n^{\prime}}\right\rangle\left\langle u_{n^{\prime}}\right|=1$. 
We only focus on the conduction band with $n=+$ (the scattering process happens for electrons in the conduction band), thus from now on we omit the band index $n$. The GH shift in the $y$ direction for a given $k_{y}$ is 
\begin{eqnarray}
\Delta_{GH}(k_{y})=\int_{-T(k_{y})}^{T(k_{y})}v(k_{y}) dt =  \int_{-T(k_{y})}^{T(k_{y})}\left[ \frac{\partial E(k_{y})}{\hbar\partial k_{y}}-\Omega_{k_{y}t} \right]dt,
\end{eqnarray}
with $T(k_{y})=\sqrt{E^{2}-\hbar^{2}v_{F}^{2}k_{y}^{2}}$ [see Fig.~\ref{s4}(c)]. From the maintext we show that $\delta I_{GH}/\delta E_{F}=\frac{e}{h}\int\frac{dk_{y}}{2\pi}\Delta_{GH}$.
Then we have
\begin{eqnarray}
\delta I_{GH}/\delta E_{F}=-\frac{e}{h}\int_{-E/\hbar v_{F}}^{E/\hbar v_{F}} dk_{y}/2\pi \int_{-T(k_{y})}^{T(k_{y})}  \Omega_{k_{y}t}dt=-\frac{e}{h}\Gamma(C)/2\pi.
\end{eqnarray} 
Since the band structure is symmetric with respect to $k_{y}$, the integration of $\partial E_{n}(k_{y})/\hbar\partial k_{y}$ vanishes.
Eq. (S36) is nothing but the Berry phase on the boundary $C$ of the integration manifold. Intuitively, $\Gamma(C)=\pi$ or $\Gamma(C)=-\pi$ because on the boundary $C$ we have $T^{2}+\hbar^{2}v_{F}^{2}k_{y}^{2}=E^{2}$ with $m=0$. In such a case, $C$ can be viewed as the Fermi surface of the massless Dirac cone, thus the Berry phase around $C$ should be $\pm\pi$. However, the sign of $\Gamma(C)$ directly determines the direction of the chiral GH shift current according to Eq. (S36). To settle down this issue, the specific form of $\Omega_{k_{y}t}$ should be given. 
	
Define $k=\sqrt{t^{2}+\hbar^{2}v_{F}^{2}k_{y}^{2}}$ and $T=k{\rm cos}\theta$, $\hbar v_{F} k_{y}=k{\rm sin}\theta$. The integral in Eq. (S36) can be performed in the polar coordinate system as
\begin{eqnarray}
\delta I_{GH}/\delta E_{F}=-\frac{e}{2\pi h}\int_{0}^{2\pi} d\theta \int_{0}^{E} kdk \Omega_{k\theta},
\end{eqnarray} 
where the Berry curvature 
\begin{eqnarray}
\Omega_{k\theta}=\frac{1}{k}i\left[ \left \langle \frac{\partial u}{\partial k}\bigg| \frac{\partial u}{\partial \theta} \right\rangle- \left \langle \frac{\partial u}{\partial \theta}\bigg| \frac{\partial u}{\partial k} \right\rangle\right].
\end{eqnarray}
Rewriting the spinor in the polar coordinate system as $|u(k,\theta)\rangle=[E+ m,-ike^{-i\theta}]^{T}/\sqrt{2E(E+ m)}$ and use the relation $k^{2}+m^{2}=E^{2}$, we obtain the expression of $\Omega_{k\theta}$ after some derivations as
\begin{eqnarray}
\Omega_{k\theta}=-\frac{1}{2Em}=-\frac{{\rm sgn}(m)}{2E\sqrt{E^{2}-k^{2}}}.
\end{eqnarray}
Therefore, the Berry phase 
\begin{eqnarray}
\Gamma(C)=\int_{0}^{2\pi} d\theta \int_{0}^{E} kdk \frac{-{\rm sgn}(m)}{2E\sqrt{E^{2}-k^{2}}}=-{\rm sgn}(m)\pi.
\end{eqnarray}
We conclude that the half-quantized chiral GH shift current $\delta I_{GH}/\delta E_{F}=\frac{e}{2h}{\rm sgn}(m)$ is protected by the $\pi$ Berry phase of the massless Dirac electron while its direction is determined by the mass $m$ of the massive barrier. Furthermore, it is quite easy to conclude that the result in Eq. (S40) is not affected by the potential $U$ appeared in Eq. (S1), since it does not affect the $\pi$ Berry phase and ${\rm sgn}(m)$. Nevertheless, the $\pi$ Berry phase may also be influenced by the finite size gap or side surface random magnetization induced gap \cite{lu_competition_2011}. In these cases, the Berry phase is $\pi(1-\frac{\delta}{E_{F}})$ where $\delta$ is the induced side surface gap. For very small $\delta$, the Berry phase is approximately $\pi$ and our analysis works well. Physically, the dependence of $\delta I_{GH}$ on $m$ can be viewed as a consequence of the time-reversal symmetry breaking.
	
We make one more discussion on the relationship between the half-quantized current and the half-quantized charge pump as investigated in \cite{xiao_berry_2010}. According to Eq. (2.6) in \cite{xiao_berry_2010}, the adiabatic charge pump is $c_{n}=-e\int_{0}^{T}dt\int_{BZ}\frac{dq}{2\pi}\Omega_{qt}^{n}$,
where $T$ denotes the period of the cyclic pump and $q$ denotes the momentum. Similarly, in our model the net charge pump during the reflection process is
\begin{eqnarray}
c=-e\int_{-E/\hbar v_{F}}^{E/\hbar v_{F}} dk_{y}/2\pi \int_{-T(k_{y})}^{T(k_{y})} dt \Omega_{k_{y}t}=-e\Gamma(C)/2\pi={\rm sgn}(m)\frac e2,
\end{eqnarray} 
indicating that the net charge pump for the reflection on one of the massive barrier is exactly half-charge $\frac e2$. When we take the other massive barrier to describe the side surface of the AI as depicted in Fig.~\ref{s3}(b) or the CI in Fig.~\ref{s2}(a), the two consecutive reflections on the two barriers make the charge pump process periodic. The total charge pumped along the $y$ direction is 0 $(1/2-1/2)$ in the AI and $e$ $(1/2+1/2)$ in the CI. The quantized $e$ charge pump indicates the existence of the quantized chiral edge channel in CIs.
\section{Derivation of the cross-section current density}\label{sec:8Derivation of the cross-section current density}
The 3D TI Hamiltonian \cite{liu_model_2010,zhang_topological_2009} used in calculating the cross-section current density and the conductances is
\begin{eqnarray}
H=\sum_{i=x,y,z} Ak_{i}\tau_{x}\otimes \sigma_{i}+(M_{0}-Bk^{2})\tau_{z}\otimes\sigma_{0}+M(\bm{r})\tau_{0}\otimes\sigma_{z},
\end{eqnarray}
where $M(\bm{r})$ represents the spatially varying magnetization the couples to the TI through the Zeeman interaction. Here, we take the system to be infinite in the $y$ direction. The cross-section in the $x-z$ plane is finite in $x$ and $y$ directions for the AI and the CI, and semi-infinite in the $x$ direction for the semi-magnetic TI. Since the Hamiltonian is infinite in $y$ direction for the three cases, $k_{y}$ is a good quantum number and the total Hamiltonian $H$ can be decomposed into summations of the momentum-sliced Hamiltonians as
\begin{eqnarray}
H=\int_{-\pi}^{\pi}\frac{dk_{y}}{2\pi}H(k_{y},\bm{r}),
\end{eqnarray}
where $H(k_{y},\bm{r})$ is the momentum-sliced Hamiltonian with momentum $k_{y}$ where $\bm{r}=(x,z)$. We define the Green's function
\begin{eqnarray}
{\mathbf G}^{r}_{k_{y}}(E)=\frac{\mathbb{I}}{E-{\mathbf H}_{k_{y}}+i0^{+}}=\sum_{n}\frac{|\psi_{k_{y},n}\rangle \langle \psi_{k_{y},n}|}{E-E_{k_{y},n}+i0^{+}},
\end{eqnarray}
where $|\psi_{n}\rangle$ is the $n_{\rm th}$ eigenstate of $H(k_{y},\bm{r})$.
Writing ${\mathbf G}^{r}_{k_{y}}(E)$ in a real space form as $G^{r}_{k_{y}}(E,{\bm r},{\bm r}')=\langle {\bm r} |{\mathbf G}^{r}_{k_{y}}(E)|{\bm r}'\rangle$.
	
The velocity operator in the $y$ direction for a given $k_{y}$ is $v_{y}(k_{y},\bm{r})=\partial H(k_{y},\bm{r})/\hbar\partial k_{y}$. The local current density in the $x-z$ plane can be expressed as
\begin{eqnarray}
J_{y}(E,\bm{r})=\int_{-\pi}^{\pi}\frac{dk_{y}}{2\pi}j_{y,k_{y}}(E,\bm{r}),
\end{eqnarray}
where
\begin{eqnarray}
j_{y,k_{y}}(E,\bm{r})=-\frac{1}{\pi}\frac{e}{\hbar}{\rm ImTr}[\frac{\partial H(k_{y})}{\partial k_{y}}G^{r}_{k_{y}}(E,\bm{r},\bm{r})]
\end{eqnarray}
is the local current density for a given $k_{y}$. Eq. (S46) can be derived as follows
\begin{eqnarray}
j_{y,k_{y}}(E,\bm{r})&&=ev(k_{y},\bm{r})\rho(E,\bm{r})=\sum_{n}ev_{n}(k_{y},\bm{r})\delta(E-E_{k_{y},n})\\
&&=-\frac{1}{\pi}{\rm Im}[\sum_{n}\mathcal{P}\frac{ev_{n}(k_{y},\bm{r})}{E-E_{k_{y},n}}-i\pi\sum_{n}ev_{n}(k_{y},\bm{r})\delta(E-E_{k_{y},n})]\\
&&=-\frac{1}{\pi}{\rm Im}[\sum_{n}\frac{\langle\psi_{k_{y},n}({\bm r})|\frac{e\partial H(k_{y})}{\hbar\partial k_{y}}|\psi_{k_{y},n}({\bm r})\rangle}{E-E_{k_{y},n}+i0^{+}}]\\
&&=-\frac{1}{\pi}\frac{e}{\hbar}{\rm ImTr}[\frac{\partial H(k_{y})}{\partial k_{y}}G^{r}_{k_{y}}(E,\bm{r},\bm{r})].
\end{eqnarray}
In Eq. (S47), $\rho(E,{\bm r})$ is the local density of states at energy $E$ and position ${\bm r}$, which can be expanded as $\rho(E,{\bm r})=\sum_{n}\delta(E-E_{k_{y},n})$. In deriving Eq. (S49) from Eq. (S48), we used the relation
\begin{eqnarray}
\frac{1}{\omega+i0^{+}}=\mathcal{P}\frac{1}{\omega}-i\pi\delta(\omega),
\end{eqnarray}
where $\mathcal{P}$ denotes the Cauchy principal value.
\begin{figure}[htbp]
	\includegraphics[width=0.8\textwidth]{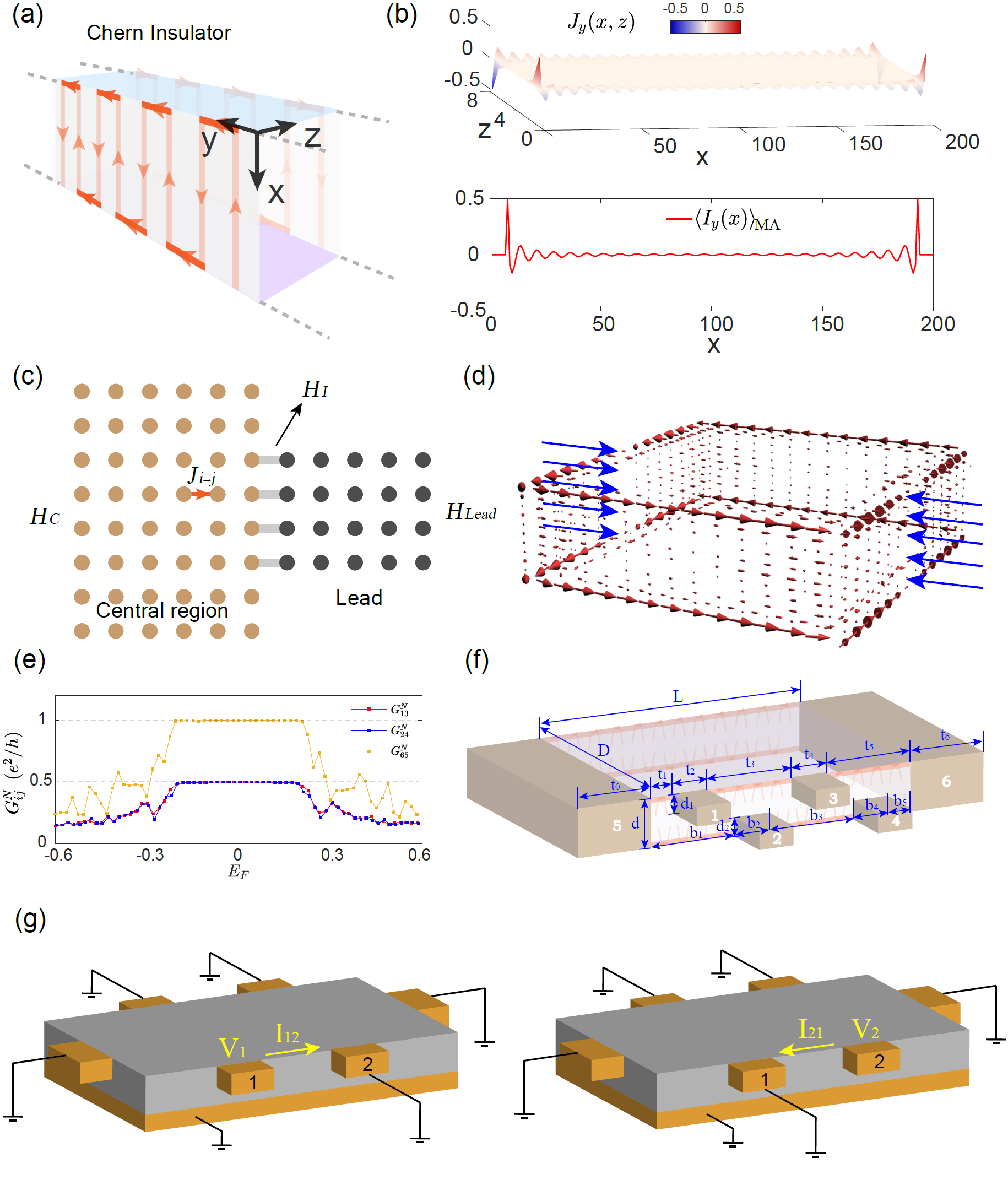}
	\caption{(a) Schematic of the side surface hinge current in the CI. (b) The upper panel shows the distribution of $J_{y}(x,z)$ in the $x-z$ plane. The lower panel shows the moving averaged flux $\langle I_{y}(\bar{x}) \rangle_{\rm MA}$ through the window $[\bar{x}-7\leqslant x\leqslant \bar{x}+7, 0\leqslant z\leqslant 4]$. Both current density $J_{y}$ and flux $I_{y}$ are in units of $e/h\Delta E$. (c) Schematic diagram showing how to calculate the local current density distributions. The central region described by the Hamiltonian $H_{C}$ is connected to a external lead $H_{Lead}$ through the coupling Hamiltonian $H_{I}$. $J_{i\to j}$ denotes the local current from site $i$ to site $j$. (d) Local current distribution of the CI. (e) The nonreciprocal conductance $G^{N}_{ij}=G_{ij}-G_{ji}$ in the CI.  (f) Schematic of the six-terminal device. Leads 5 and 6 connect to the ends of the CI film. Terminals 1$\sim$4 are surface leads with leads 1 and 3 (2 and 4) connected to the top (bottom) surface of the sample. (f) Schematic of the six-terminal device and the illustration of the relationship between the quantized chiral edge transport and the half-quantized hinge transport in the CI phase. (g) Schematic of the experimental setup to measure the nonreciprocal conductances $G^{N}_{ij}$ between leads. When measuring $G_{12}$, the voltage $V_{1}$ is applied to lead 1 with all the other leads grounded and collect the current $I_{12}$ flowing into lead 2. The conductance is calculated through $G_{12}=I_{12}/V_{1}$. Then measure $G_{21}=I_{21}/V_{2}$. The difference between $G_{12}$ and $G_{21}$ gives $G^{N}_{12}$.}  
	\label{s5}
\end{figure}	
\section{Derivation of the differential conductance and the local current density}\label{sec:9Derivation of the differential conductance and the local current density}
In this section, we derive the differential conductance and the local current density with the help of the non-equilibrium Green's function method \cite{haug_hartmut_and_jauho_antti-pekka_and_others_quantum_2008,mackinnon_calculation_1985,metalidis_greens_2005}. As depicted in Fig.~\ref{s5}(c), we first consider the simplest case where the central region connects to an external lead. $H_{C}$ and $H_{Lead}$ are the Hamiltonians of the central region and the lead, $H_{I}$ is the coupling between the central region and the lead. We use $a_{{\mathbf i}'}$ to denote the annihilation operator in the central region at site ${\mathbf i}'$ and $c_{{\mathbf i}}$ to denote the annihilation operator in the lead at site ${\mathbf i}$. The coupling Hamiltonian reads
\begin{eqnarray}
H_{I}=t_{{\mathbf i}{\mathbf i}'}c^{\dagger}_{{\mathbf i}}a_{{\mathbf i}'}+t^{*}_{{\mathbf i}'{\mathbf i}}a^{\dagger}_{{\mathbf i}'}c_{{\mathbf i}}.
\end{eqnarray}
The particle leakage on the lead is
\begin{eqnarray}
-\sum_{{\mathbf i}}\frac{dN_{{\mathbf i}}}{dt}&&=-\sum_{{\mathbf i}}\frac{d}{dt}\langle c^{\dagger}_{{\mathbf i}}c_{{\mathbf i}}\rangle=-\sum_{{\mathbf i}}\frac{1}{i\hbar}\langle [c^{\dagger}_{{\mathbf i}}c_{{\mathbf i}}, H]\rangle\\
&&=-\sum_{{\mathbf i}{\mathbf i}'}\frac{1}{i\hbar}\langle t_{{\mathbf i}{\mathbf i}'}c^{\dagger}_{{\mathbf i}}a_{{\mathbf i}'}+t^{*}_{{\mathbf i}'{\mathbf i}}a^{\dagger}_{{\mathbf i}'}c_{{\mathbf i}}\rangle,
\end{eqnarray}
where $H=H_{C}+H_{Lead}+H_{I}$. From Eq. (S53) to Eq. (S54) we used the relation $[H_{C},\sum_{{\mathbf i}}N_{{\mathbf i}}]=[H_{Lead},\sum_{{\mathbf i}}N_{{\mathbf i}}]=0$.
Through relations
\begin{eqnarray}
\nonumber
i\langle c^{\dagger}_{{\mathbf i}}(t=0)a_{{\mathbf i}'}(t=0) \rangle&&=G^{<}_{{\mathbf i}'{\mathbf i}}(t=0)=\int_{-\infty}^{\infty}\frac{dE}{2\pi}G^{<}_{{\mathbf i}'{\mathbf i}}(E)\\
i\langle a^{\dagger}_{{\mathbf i}'}(t=0)c_{{\mathbf i}}(t=0) \rangle&&=G^{<}_{{\mathbf i}{\mathbf i}'}(t=0)=\int_{-\infty}^{\infty}\frac{dE}{2\pi}G^{<}_{{\mathbf i}{\mathbf i}'}(E).
\end{eqnarray}
Combine Eq. (S54) and Eq. (S55), we obtain the net current flowing into the lead as
\begin{eqnarray}
I_{Lead}=-e\sum_{{\mathbf i}}\frac{dN}{dt}=\frac{e}{\hbar}\sum_{{\mathbf i}{\mathbf i}'}\int_{-\infty}^{\infty}\frac{dE}{2\pi}[t_{{\mathbf i}{\mathbf i}'}G^{<}_{{\mathbf i}'{\mathbf i}}(E)-t^{*}_{{\mathbf i}'{\mathbf i}}G^{<}_{{\mathbf i}{\mathbf i}'}(E)].
\end{eqnarray}
We denote the Green's function of the free lead (without coupling to the central region) as ${\mathbf g}$, then use the Langreth theorem and Dyson equations we have
\begin{eqnarray}
G^{<}_{{\mathbf i}'{\mathbf i}}&&=\sum_{{\mathbf j}{\mathbf j}'}(G_{{\mathbf i}'{\mathbf j}'}t^{*}_{{\mathbf j}'{\mathbf j}}g_{{\mathbf j}{\mathbf i}})^{<}=\sum_{{\mathbf j}{\mathbf j}'}[G^{r}_{{\mathbf i}'{\mathbf j}'}t^{*}_{{\mathbf j}'{\mathbf j}}g^{<}_{{\mathbf j}{\mathbf i}}+G^{<}_{{\mathbf i}'{\mathbf j}'}t^{*}_{{\mathbf j}'{\mathbf j}}g^{a}_{{\mathbf j}{\mathbf i}}]\\
G^{<}_{{\mathbf i}{\mathbf i}'}&&=\sum_{{\mathbf j}{\mathbf j}'}(g_{{\mathbf i}{\mathbf j}}t^{*}_{{\mathbf j}{\mathbf j}'}G_{{\mathbf j}'{\mathbf i}'})^{<}=\sum_{{\mathbf j}{\mathbf j}'}[g^{r}_{{\mathbf i}{\mathbf j}}t^{*}_{{\mathbf j}{\mathbf j}'}G^{<}_{{\mathbf j}'{\mathbf i}'}+g^{<}_{{\mathbf i}{\mathbf j}}t^{*}_{{\mathbf j}{\mathbf j}'}G^{a}_{{\mathbf j}'{\mathbf i}'}].
\end{eqnarray}
Then we have
\begin{eqnarray}
I_{Lead}=\frac{e}{\hbar}\sum_{{\mathbf i}{\mathbf i}'{\mathbf j}{\mathbf j}'}\int_{-\infty}^{\infty}\frac{dE}{2\pi}(t_{{\mathbf i}{\mathbf i}'}G^{r}_{{\mathbf i}'{\mathbf j}'}t^{*}_{{\mathbf j}'{\mathbf j}}g^{<}_{{\mathbf j}{\mathbf i}}+t_{{\mathbf i}{\mathbf i}'}G^{<}_{{\mathbf i}'{\mathbf j}'}t^{*}_{{\mathbf j}'{\mathbf j}}g^{a}_{{\mathbf j}{\mathbf i}}-t_{{\mathbf i}'{\mathbf i}}g^{r}_{{\mathbf i}{\mathbf j}}t^{*}_{{\mathbf j}{\mathbf j}'}G^{<}_{{\mathbf j}'{\mathbf i}'}-t_{{\mathbf i}'{\mathbf i}}g^{<}_{{\mathbf i}{\mathbf j}}t^{*}_{{\mathbf j}{\mathbf j}'}G^{a}_{{\mathbf j}'{\mathbf i}'}).
\end{eqnarray}
We define self-energies
\begin{eqnarray}
\Sigma_{{\mathbf j}'{\mathbf i}'}^{</r/a}&&=\sum_{{\mathbf i}{\mathbf j}}t^{*}_{{\mathbf j}'{\mathbf j}}g^{</r/a}_{{\mathbf j}{\mathbf i}}t_{{\mathbf i}{\mathbf i}'},
\end{eqnarray}
and substitute it into Eq. (S59) as 
\begin{eqnarray}
\nonumber
I_{Lead}&&=\frac{e}{\hbar}\int_{-\infty}^{\infty}\frac{dE}{2\pi}(G^{r}_{{\mathbf i}'{\mathbf j}'}\Sigma^{<}_{{\mathbf j}'{\mathbf i}'}+G^{<}_{{\mathbf i}'{\mathbf j}'}\Sigma^{a}_{{\mathbf j}'{\mathbf i}'}-G^{<}_{{\mathbf j}'{\mathbf i}'}\Sigma^{r}_{{\mathbf i}'{\mathbf j}'}-G^{a}_{{\mathbf j}'{\mathbf i}'}\Sigma^{<}_{{\mathbf i}'{\mathbf j}'})\\
&&=\frac{e}{\hbar}\int_{-\infty}^{\infty}\frac{dE}{2\pi}[\Sigma^{<}_{{\mathbf i}'{\mathbf j}'}(G^{r}_{{\mathbf j}'{\mathbf i}'}-G^{a}_{{\mathbf j}'{\mathbf i}'})+G^{<}_{{\mathbf i}'{\mathbf j}'}(\Sigma^{a}_{{\mathbf j}'{\mathbf i}'}-\Sigma^{r}_{{\mathbf j}'{\mathbf i}'})].
\end{eqnarray}
According to the fluctuation-dissipation theorem
\begin{eqnarray}
g^{<}_{{\mathbf j}{\mathbf i}}=if_{Lead}A^{<}_{{\mathbf j}{\mathbf i}}=-f_{Lead}(g^{r}_{{\mathbf j}{\mathbf i}}-g^{a}_{{\mathbf j}{\mathbf i}}),
\end{eqnarray}
where ${\mathbf A}^{<}=i({\mathbf g}^{r}-{\mathbf g}^{a})$ is the spectral function and $f_{Lead}=1/(e^{(E-\mu_{n})/k_{B}T}+1)$ is the Fermi distribution function of the lead. We then have
\begin{eqnarray}
\Sigma^{<}_{{\mathbf i}'{\mathbf j}'}=-f_{Lead}[\Sigma^{r}_{{\mathbf i}'{\mathbf j}'}-\Sigma^{a}_{{\mathbf i}'{\mathbf j}'}].
\end{eqnarray}
Define the linewidth function $\Gamma_{Lead,{\mathbf i}'{\mathbf j}'}=i(\Sigma^{r}_{{\mathbf i}'{\mathbf j}'}-\Sigma^{a}_{{\mathbf i}'{\mathbf j}'})$ and substitute it into Eq. (S61)
\begin{eqnarray}
\nonumber
I_{Lead}&&=\frac{e}{\hbar}\int_{-\infty}^{\infty}\frac{dE}{2\pi}[i\Gamma_{Lead,{\mathbf i}'{\mathbf j}'}f_{Lead}(G^{r}_{{\mathbf j}'{\mathbf i}'}-G^{a}_{{\mathbf j}'{\mathbf i}'})+i\Gamma_{Lead,{\mathbf i}'{\mathbf j}'}G^{<}_{{\mathbf i}'{\mathbf j}'}]\\
&&=\frac{ie}{\hbar}\int_{-\infty}^{\infty}\frac{dE}{2\pi}{\rm Tr}[f_{Lead}{\mathbf \Gamma}_{Lead}({\mathbf G}^{r}-{\mathbf G}^{a})+{\mathbf \Gamma}_{Lead}{\mathbf G}^{<}].
\end{eqnarray}
	
Now we consider a multi-terminal system. We use ${\mathbf \Gamma}_{n}$ and $f_{n}$ to denote the linewidth function and the Fermi distribution function of the $n_{\rm th}$ lead. Define the total self-energy from all the leads 
\begin{eqnarray}
{\mathbf \Sigma}^{<}=\sum_{n}-f_{n}({\mathbf \Sigma}^{r}-{\mathbf \Sigma}^{a})=\sum_{n}if_{n}{\mathbf \Gamma}_{n}.
\end{eqnarray}
According to the Keldysh formula we have 
\begin{eqnarray}
{\mathbf G}^{<}={\mathbf G}^{r}{\mathbf \Sigma}^{<}{\mathbf G}^{a}=\sum_{n}{\mathbf G}^{r}if_{n}{\mathbf \Gamma}_{n}{\mathbf G}^{a}.
\end{eqnarray}
The Dyson equations ${\mathbf G}^{r/a}={\mathbf g}^{r/a}+{\mathbf g}^{r/a}{\mathbf \Sigma}^{r/a}{\mathbf G}^{r/a}$ implies $[{\mathbf g}^{r/a}]^{-1}=[{\mathbf G}^{r/a}]^{-1}+{\mathbf \Sigma}^{r/a}$. Note that $[{\mathbf g}^{r}]^{-1}=[{\mathbf g}^{a}]^{-1}=E-{\mathbf H}_{Lead}$, therefore $[{\mathbf G}^{a}]^{-1}-[{\mathbf G}^{r}]^{-1}={\mathbf \Sigma}^{r}-{\mathbf \Sigma}^{a}=\sum_{n}-i{\mathbf \Gamma}_{n}$. Then we have 
\begin{eqnarray}
{\mathbf G}^{r}-{\mathbf G}^{a}=\sum_{n}-i{\mathbf G}^{r}{\mathbf \Gamma}_{n}{\mathbf G}^{a}.
\end{eqnarray}
Combine Eq. (S64), Eq. (S66), and Eq. (S67), we obtain the current flowing into the $m_{\rm th}$ terminal as
\begin{eqnarray}
\nonumber
I_{m}&&=\frac{e}{\hbar}\int_{-\infty}^{\infty}\frac{dE}{2\pi}\sum_{n}{\rm Tr}[{\mathbf \Gamma}_{m}f_{m}{\mathbf G}^{r}{\mathbf \Gamma}_{n}{\mathbf G}^{a}-{\mathbf \Gamma}_{m}{\mathbf G}^{r}f_{n}{\mathbf \Gamma}_{n}{\mathbf G}^{a}]\\
&&=\frac{e}{\hbar}\int_{-\infty}^{\infty}\frac{dE}{2\pi}\sum_{n}(f_{m}-f_{n}){\rm Tr}[{\mathbf \Gamma}_{m}{\mathbf G}^{r}{\mathbf \Gamma}_{n}{\mathbf G}^{a}].
\end{eqnarray}
At zero temperature, the Fermi distribution function becomes $f_{n}(E)=\Theta(E_{F,n}-E)=\Theta(-eV_{n}-E)$, where $\Theta(x)$ is the Heaviside step function and $V_{n}$ is the gate voltage applied to the terminal $n$. For small bias we take the approximation $\int_{-\infty}^{\infty}dE(f_{m}-f_{n})(\cdot\cdot\cdot)\approx (E_{F,m}-E_{F,n})(\cdot\cdot\cdot)=e(V_{n}-V_{m})(\cdot\cdot\cdot)$ and we have
\begin{eqnarray}
I_{m}=\frac{e^{2}}{h}\sum_{n}(V_{n}-V_{m})T_{mn}=G_{mn}(V_{m}-V_{n}),
\end{eqnarray}
where the differential conductance $G_{mn}=\frac{e^{2}}{h}T_{mn}$ and the transmission coefficient $T_{mn}={\rm Tr}[{\mathbf \Gamma}_{m}{\mathbf G}^{r}{\mathbf \Gamma}_{n}{\mathbf G}^{a}]$.
	
We now derive the local current density distribution \cite{haug_hartmut_and_jauho_antti-pekka_and_others_quantum_2008,jiang_numerical_2009,mackinnon_calculation_1985,metalidis_greens_2005}. Consider the local Hamiltonian on a given site ${\mathbf i}$ as 
\begin{eqnarray}
H_{I}=\sum_{{\mathbf j}}(H_{{\mathbf j}{\mathbf i}}c_{{\mathbf j}}^{\dagger}c_{{\mathbf i}}+H_{{\mathbf i}{\mathbf j}}c_{{\mathbf i}}^{\dagger}c_{{\mathbf j}}).
\end{eqnarray}
The current flowing into the site ${\mathbf i}$ is
\begin{eqnarray}
\nonumber
I_{{\mathbf i}}&&=-e\langle \dot{N}_{\mathbf i}\rangle=-\frac{e}{i\hbar}\langle [N_{\mathbf i},H_{I}] \rangle\\\nonumber
&&=-\frac{e}{i\hbar}\langle [c_{{\mathbf i}}^{\dagger}c_{{\mathbf i}},\sum_{{\mathbf j}}H_{{\mathbf j}{\mathbf i}}c_{{\mathbf j}}^{\dagger}c_{{\mathbf i}}+H_{{\mathbf i}{\mathbf j}}c_{{\mathbf i}}^{\dagger}c_{{\mathbf j}}] \rangle\\ \nonumber
&&=-\frac{e}{i\hbar}\langle \sum_{{\mathbf j}}-H_{{\mathbf j}{\mathbf i}}c_{{\mathbf j}}^{\dagger}c_{{\mathbf i}}+H_{{\mathbf i}{\mathbf j}}c_{{\mathbf i}}^{\dagger}c_{{\mathbf j}} \rangle \\
&&=\sum_{{\mathbf j}} J_{{\mathbf j}\to{\mathbf i}}.
\end{eqnarray}
The local current from site ${\mathbf j}$ to ${\mathbf i}$ [see Fig.~\ref{s5}(c)] can further be expressed as
\begin{eqnarray}
\nonumber
J_{{\mathbf j}\to{\mathbf i}}&&=-\frac{e}{i\hbar}(-H_{{\mathbf j}{\mathbf i}}\langle c_{{\mathbf i}}^{\dagger}c_{{\mathbf i}} \rangle+H_{{\mathbf i}{\mathbf j}}\langle c_{{\mathbf i}}^{\dagger}c_{{\mathbf j}} \rangle)\\ \nonumber
&&=-\frac{e}{i\hbar}(iH_{{\mathbf j}{\mathbf i}}G^{<}_{{\mathbf i}{\mathbf j}}-iH_{{\mathbf i}{\mathbf j}}G^{<}_{{\mathbf j}{\mathbf i}})\\
&&=-\frac{2e}{\hbar}\int_{-\infty}^{\infty}\frac{dE}{2\pi}{\rm ReTr}[H_{{\mathbf j}{\mathbf i}}G^{<}_{{\mathbf i}{\mathbf j}}(E)].
\end{eqnarray}
With the help of Eq. (S65) and Eq. (S66), the above expression can be written as
\begin{eqnarray}
\nonumber
J_{{\mathbf j}\to{\mathbf i}}&&=-\frac{2e}{\hbar}\int_{-\infty}^{\infty}\frac{dE}{2\pi}{\rm ReTr}[H_{{\mathbf j}{\mathbf i}}\sum_{n}G^{r}_{{\mathbf i}{\mathbf i}'}(E)\Gamma_{n,{\mathbf i}'{\mathbf j}}(E)f_{n}G^{a}_{{\mathbf j}'{\mathbf j}}(E)]\\
&&=-\frac{2e}{\hbar}\int_{-\infty}^{\infty}\frac{dE}{2\pi}\sum_{n}{\rm ImTr}[H_{{\mathbf j}{\mathbf i}}({\mathbf G}^{r}(E)f_{n}{\mathbf \Gamma}_{n}(E){\mathbf G}^{a}(E))_{{\mathbf i}{\mathbf j}}].
\end{eqnarray}
At zero temperature, the Fermi distribution function becomes $f_{n}(E)=\Theta(E_{F,n}-E)=\Theta(-eV_{n}-E)$ and the integral becomes $\int_{-\infty}^{\infty}dE f_{n}(\cdot\cdot\cdot)=\int_{-\infty}^{-eV_{n}}dE(\cdot\cdot\cdot)$. Then it is straightforward to obtain
\begin{eqnarray}
\nonumber
J_{{\mathbf j}\to{\mathbf i}}&&=-\frac{2e}{h}\sum_{n}\int_{-\infty}^{-eV_{n}}dE{\rm ImTr}[H_{{\mathbf j}{\mathbf i}}({\mathbf G}^{r}(E){\mathbf \Gamma}_{n}(E){\mathbf G}^{a}(E))_{{\mathbf i}{\mathbf j}}]\\
&&=-\frac{2e}{h}\sum_{n}\int_{-\infty}^{0}dE{\rm ImTr}[H_{{\mathbf j}{\mathbf i}}({\mathbf G}^{r}(E){\mathbf \Gamma}_{n}(E){\mathbf G}^{a}(E))_{{\mathbf i}{\mathbf j}}]-\frac{2e}{h}\sum_{n}\int_{0}^{-eV_{n}}dE{\rm ImTr}[H_{{\mathbf j}{\mathbf i}}({\mathbf G}^{r}(E){\mathbf \Gamma}_{n}(E){\mathbf G}^{a}(E))_{{\mathbf i}{\mathbf j}}].
\end{eqnarray}
The first term in Eq. (S74) represents the equilibrium current while the second term represents the non-equilibrium transport current. For small bias, the non-equilibrium transport current can be simplified as
\begin{eqnarray}
J_{neq,{\mathbf j}\to{\mathbf i}}=-\frac{2e}{h}\sum_{n}V_{n}{\rm ImTr}[H_{{\mathbf j}{\mathbf i}}({\mathbf G}^{r}(E){\mathbf \Gamma}_{n}(E){\mathbf G}^{a}(E))_{{\mathbf i}{\mathbf j}}].
\end{eqnarray}
\section{Chiral edge transport and its relation to half-quantized hinge channels in Chern insulators}\label{sec:10Chiral edge transport and its relation to half-quantized hinge channels in Chern insulators}
	
The cross-section local current density for the CI is shown in Fig.~\ref{s5}(a) and (b). When side surface electrons bounce back and forth between the top and bottom surfaces, the direction of the GH shift currents on the two hinges flow in the same direction, giving rise to chiral net side surface current. Fig.~\ref{s5}(b) shows the spatial distribution of the cross-section local current density for the CI. Similar to the AI case as analyzed in the main text, the local current peaks at the hinges, but the hinge currents flow in the same direction, leading to a net chiral side surface current which is in sharp distinction from the AI. In the upper panel of Fig.~\ref{s5}(d) we also plot the local transport current distribution calculated by Eq. (S75). Here, the CI bulk is connected to two external leads as depicted by the blue arrows in the upper panel of Fig.~\ref{s5}(d) and we take $V_{1}=V_{2}$ to investigate the chiral or helical nature of the side surface current. In Fig.~\ref{s5}(e) we demonstrate that the nonreciprocal conductances on the top and bottom hinges of the CI have the same sign, contributing to totally quantized side surface transport. In Fig.~\ref{s5}(f) we illustrate that the quantized chiral conductance channel in the CI originates from the combination of the two half-quantized hinge channels (1/2+1/2).
\section{Experimental setups to measure the nonreciprocal conductances}\label{sec:11Experimental setups to measure the nonreciprocal conductances}
In this section, we illustrate the principles in measuring the nonreciprocal conductances in multi-terminal devices. We consider the nonreciprocal conductances between two external leads [such as lead 1 and lead 2 as shown in Fig.~\ref{s5}(g)]. To obtain $G^{N}_{12}$, the conductances $G_{12}$ and $G_{21}$ should be measured. $G_{12}$ is defined as $G_{12}=I_{12}/V_{1}$, where $V_{1}$ is the applied gate voltage on lead 1 with all the other leads grounded as shown in Fig.~\ref{s5}(g), and $I_{12}$ is the current flowing into lead 2. Here, we only consider the differential conductances so that the $I_{12}$ is the differential current induced by the small bias $V_{1}$. The measurement of $G_{21}=I_{21}/V_{2}$ is similar. 

In experiments, fabricating the multi-terminal device in Fig.~\ref{s5}(g) (the bottom surface of the sample is grounded and all the other leads are connected near the top surface of the sample) may be easier than the six-terminal device shown in Fig.~\ref{s5}(f) \cite{zhang_non-reciprocal_2022}. Moreover, since the bottom surface of the sample is grounded in Fig.~\ref{s5}(f), the half-quantized hinge channel localized on the bottom hinge of the AI or CI is inactive, thus only the half-quantized hinge channel on the top hinge contributes to the nonreciprocal conductance. Therefore, the measurement result of $G^{N}_{12}$ is sensitive to the quality of lead 1 and lead 2. To improve the experimental accuracy, we emphasize that the surface leads [lead 1 and lead 2 in Fig.~\ref{s5}(f)] should be thick enough (but do not touch the bottom surface) such that they couple to more conducting side surface channels. Besides, lead 1 and lead 2 should be close enough to ensure that nearly all hinge current can flow into the measuring lead. 
\section{Model parameters in numerics}\label{sec:12Model parameters in numerics}
In \ref{sec:4Anomalous velocity and band modification by the GH shift}, the parameters in the Hamiltonian Eq. (S16) are $\hbar v_{F}/a=1$. For the gapless region $\mu=0.2$ and $m=0$, for the gapped region $\mu=0$ and $m=0.03$. $L_{x}=600$.
	
In \ref{sec:10Chiral edge transport and its relation to half-quantized hinge channels in Chern insulators} and the main text, in calculating the cross-section local current density, we take the parameters of the Hamiltonian Eq. (S42) as $A=1$, $B=0.6$, $M_{0}=1$, $E_{F}=0.4$ and $M=0.6$. In the semi-magnetic TI case, the magnetization term $H_{M}=0.6\times\tau_{0}\otimes\sigma_{z}$ only couples to the top surface, while in the AI/CI case $H_{M}=\pm0.6\times\tau_{0}\otimes\sigma_{z}$ couple to the top and bottom surfaces.
	
In calculating the local current distributions we take $A=1$, $B=0.6$, $M_{0}=1$, $E_{F}=0.4$ and $M=0.6$. The system size (for both AI and CI) is $15\times15\times30$. In plotting Fig.~\ref{s5}(d) and the Fig.~4(b) in the main text, the thickness is squeezed but does not affect the results in demonstrating the helical/chiral transport nature on the AI/CI side surface.
	
In calculating the nonreciprocal conductances, we take $A=1$, $B=0.6$, $M_{0}=1$, $E_{F}=0.4$ and $M=0.2$. The geometrical size of the six-terminal device is marked in Fig.~\ref{s5}(e). We take $t_{0}=25$, $t_{1}=1$, $t_{2}=11$, $t_{3}=1$, $t_{4}=11$, $t_{5}=7$, $t_{6}=25$, $b_{1}=7$, $b_{2}=11$, $b_{3}=1$, $b_{4}=11$, $b_{4}=1$, $d_{1}=10$, $d_{2}=10$, $d=21$, $D=20$, and $L=31$.   
\bibliography{GH}